# Generalized plasmon-pole approximation for the dielectric function of a multicomponent plasma

**B J B Crowley**[1,2,3]

[1]Department of Physics, University of Oxford, Parks Road, Oxford OX1 3PU, UK
[2]AWE PLC, Reading RG7 4PR, UK

[3]Email: *basil.crowley@physics.ox.ac.uk*

This article reviews two currently available analytic models of the dielectric function of a plasma consisting of quantum particles interacting via Coulomb forces, namely the Random Phase Approximation (RPA) and the Standard (Simple) Plasmon Pole Approximation (SPPA). It is shown that these models describe different non-overlapping plasma regimes. The RPA describes a weakly coupled plasma whose dynamics are an admixture of independent particle modes at short wavelengths and collective plasma modes at long wavelengths. The SPPA describes a system dominated by collective resonances at all wavelengths. A new model of a multicomponent plasma, the Generalized Plasmon Pole Approximation (GPPA), is formulated and shown to provide a description of a range of plasma regimes encompassing both those described by the RPA and the SPPA. Therefore, as well as providing a good representation of the RPA, in both quantum and classical, regimes, this new model also provides a promising basis for modelling non-ideal quantum plasmas, which are plasmas in regimes of significantly stronger coupling than are addressed by the RPA.

This page is intentionally left blank

# CONTENTS

# 1 SUMMARY

## 1.1 Introduction

Analytical models of the dielectric function of Coulomb many-body systems are a potentially powerful tool in the modelling of plasmas and metals across a wide range of conditions. The advantages of such methods are that they are holistic in that they automatically span all scales of length and time; in which sense, they possess a self-consistency that ensures that essential relationships, such as causality and various sum-rules, between different properties of a material are maintained. At non-relativistic energies, the constituent particles, which make up ordinary matter (atomic nuclei and electrons, or, at a less fundamental level, ions and electrons) can generally be considered to interact solely via Coulomb forces, representing the fundamental electromagnetic interaction between quasi-stationary charges. The motion of such charges also generates and creates an interaction with magnetic fields, but, in low-temperature non-relativistic systems, these forces are much weaker that the static Coulomb forces. In very hot relativistic plasmas, and deep inside heavy atoms, the electrodynamics must be treated relativistically and magnetic and electrostatic fields are treated together. Analytical methods are particularly useful for treating the collective modes of a many particle system, such modes being typically represented by complex zeros or singularities of Green functions.

The purpose of this article is to describe some analytical methods for treating the dielectric function of Coulomb plasmas in regimes of weak coupling (ideal plasmas) and strong coupling (solid-state plasmas) and to introduce a new analytical model that spans all regimes.

## 1.2 Dielectric Formalism

The dielectric function describes the response of a material system in equilibrium to (weak) electric fields, where "weak" essentially means that non-linear terms in the response can be neglected.

For a system consisting solely of particles interacting via Coulomb forces, the dielectric function is a representation of the many particle Green function. In principle, complete knowledge of the dielectric function provides information about all of the equilibrium properties of the system. In practice, this requires knowing the function $\varepsilon(\mathbf{q},\omega)$ on all length (wavenumber $q$) and temporal (frequency $\omega$) scales. Empirically, experiments only allow one to access small pieces of this function directly. This means that models are required to extrapolate the observational data to other regimes, whereby other properties of the system can be determined consistently with the phenomenological data, as well as with the theoretical constraints such as causality, sum-rules etc. Analytical methods are an important aid to achieving this.

Theoretically, the calculation of this function from first principles is not feasible, even for the simplest of physical systems, except in certain limiting cases. Coulomb forces account for most of the properties of ordinary matter, so a completely general theoretical formulation would therefore have to encompass virtually all normal states of matter, and would include information about atomic structure, molecular structure, molecular interactions, solid and liquid-state physics, phase transitions, plasma properties, which would be quite an achievement. In fact any state of matter in which the interactions between the component particles are dominated by Coulomb forces would be embodied in such a general function. While such general formalisms exist, they do not lead to tractable results, except in very idealized situations, with often quite severe limitations. Models of the dielectric function that interpolate sensibly between these idealized limits therefore provide potentially useful means of treating real systems.

The dielectric function does not include a description of magnetic forces (for which one would need the equivalent magnetic permeability function). However magnetic forces are relatively weak in plasmas where the particle velocities are non-relativistic, and can be generally ignored in the determination of the bulk properties, for example, of such systems.

Because the dielectric function is equivalent to a physical Green function, it possesses a number of important mathematical properties. The most important of these is the Landau spectral representation, according to which, any Green function, one satisfying causality, or any physically reasonable dielectric function, can be represented as a function of complex frequency $\omega$ that is analytic (free of singularities) in $\mathrm{Im}\,\omega > 0$. This is the essential property of a retarded Green function, one in which cause precedes effect. The advanced Green function, which is a corresponding function in which the singularities occur only in $\mathrm{Im}\,\omega > 0$, and which describes ‘time reversed’ processes in which effect precedes cause, also has an essential role in some theories.

The inverse dielectric function is a retarded Green function, and therefore the dielectric function has zeros only in $\mathrm{Im}\,\omega < 0$. Perhaps less obviously, the dielectric function itself is also a retarded Green function, and therefore has singularities only in $\mathrm{Im}\,\omega < 0$. This effectively means that the zeros and poles of both functions are confined to $\mathrm{Im}\,\omega < 0$.

The spectral representation also provides limiting forms at high frequencies ($\omega \to \infty$) and important relations between the real and imaginary parts of the dielectric function and its inverse. These are the well-known Kramers-Krönig dispersion relations.

The analytic properties also imply sum rules whereby the limiting properties of $\varepsilon(\mathbf{q},\omega)$ near $\omega = 0$ and $\omega = \infty$ are related to integrals of $\varepsilon(\mathbf{q},\omega)$ over frequency. Formal derivations of some of these sum rules are presented in APPENDIX A. Higher order sum rules (not derived here) known as subtracted sum rules can be derived in a similar fashion, by subtracting out the lower order behaviours, and reapplying the methodology.

The zeros of the dielectric function, which, as we have noted, are confined $\mathrm{Im}\,\omega < 0$, correspond to resonances (eg collective modes, atomic transitions). Resonance damping and line broadening are accounted for by the imaginary part of the resonant frequency.

The analytic properties of the dielectric function, combined with a knowledge of the resonant and collective behaviour able to affect the phenomena that one wishes to describe, can enable one to construct models of the dielectric functions in which these properties are incorporated in a reasonable and tractable way. Because such models automatically provide analytic continuations throughout the whole frequency domain they can potentially provide a sufficiently complete representation of the physical system to enable the deduction of the accessible properties of that system.

Three such models are described in this article. These are the Random Phase Approximation (RPA), the Simple Plasmon Pole Approximation (SPPA), which are known, and the Generalized Plasmon Pole Approximation (GPPA), which is new.

## 1.3 Random Phase Approximation

The Random Phase Approximation (RPA) [1],[2],[3] describes weakly coupled Coulomb systems, in which short range correlations between individual particles are completely neglected. In this sense, the plasma is modelled as being essentially collisionless.

The RPA incorporates a full description of quantum statistical correlations and incorporates an exact description of the finite temperature ideal Fermi gas.

In the asymptotic limit of small $q$ (long wavelength scales) the RPA describes ideal plasmas in which quantum statistics (degeneracy) is properly accounted for. In the classical limit, the RPA is equivalent to Debye-Hückel theory.

The RPA predicts and describes long wavelength collective behaviour, which takes the form of plasma oscillations. Because it is a quantum mechanical model, these excitations are represented as Bose-Einstein quasi-particles (plasmons).

However, according to the RPA, the collective modes and the independent particle translational motion are not decoupled, but are mixed. This gives rise to damping, known as Landau damping, of the collective modes, even though the plasma is deemed to be collisionless.

## 1.4 Simple Plasmon Pole Approximation

This model represents the system purely in terms of collective modes. The dielectric function is just the analytic continuation of a simple pole. In this respect, the SPPA is the simplest non-trivial model of an analytic dielectric function that one can construct.

Because the SPPA contains nothing that would correspond to independent particle translational motion, it does not encompass the RPA. The SPPA and the RPA therefore describe different types of physical system. The SPPA describes strongly coupled systems in which the dynamics are entirely due to collective behaviour, while the RPA describes weakly-coupled systems in which short-range correlations are negligible.

An example of a physical system that is dominated by collective modes is a crystalline solid, in which the collective modes are phonons. The remarkable quantitative similarities between the ion-acoustic wave velocities $C_{\mathrm{ia}} = \Omega_{\mathrm{i}} D_{\mathrm{e}} = \sqrt{\tfrac{2}{3} Z T_{\mathrm{F}} / m_{\mathrm{i}}}$ and the measured sound velocities $C_{\mathrm{s}}$ for liquid metals [4], particularly those for which the valence $Z$ is well-defined, is an indication that a simple plasmon-pole is applicable to some strongly-coupled solid-state plasmas.

However, plasmas in which the component particles have mobility, including regimes intermediate between the collectively dominated one and the RPA, would not be described.

## 1.5 Generalized Plasmon Pole Approximation

This is a new model, introduced in this article, that incorporates features of both the RPA and the SPPA.

The GPPA dielectric function encompasses the RPA, Debye-Hückel and SPPA in its various limits.

The formulation covers multicomponent plasmas at the outset.

The spatial dependence of the GPPA is defined parametrically through unspecified (but physically constrained) functions $\alpha(\mathbf{q})$, $\beta(\mathbf{q})$ and $\lambda(\mathbf{q})$ as well as the spectrum $\hat{\Omega}_{\mathbf{q}} - \tfrac{1}{2}\mathrm{i}\nu_{\mathbf{q}}$. These functions can be determined from the RPA or by whatever alternative model is deemed appropriate.

The GPPA is capable of covering all regimes intermediate between RPA and SPPA.

## 1.6 Structure Factor

The Fourier transform, in both space and time, of the two body charge-density correlation function is known as the charge-density dynamic structure factor, $S(\mathbf{q},\omega)$ [5],[3],[6]. This function therefore provides information about the charge density fluctuations in the system.

This function is directly related to the imaginary part of the reciprocal of the dielectric function by the fluctuation-dissipation theorem, so called because it directly relates the fluctuation correlation function to the dissipation (damping) implied by the imaginary part of the dielectric function.

The Static Structure Factor describes only the instantaneous spatial correlations (ie at a fixed time) and is directly related to the pair correlation function. It is relevant to interference between photons or fast particles scattered by spatially separated charges.

Thomson scattering is an example of this. However Thomson scattering effectively sees only the electrons. Any scattering by other charges (ions and nuclei) is highly suppressed due to their significantly greater masses. In order to calculate Thomson scattering, it is therefore appropriate to separate out just the electron-electron correlation part of the charge-density structure factor. On the other hand, interference between photons resonantly scattered by different atoms is dependent on the ion-ion structure factor, eg [7], while the complete description of electron-electron correlations involves a part mediated by ion-ion and electron-ion correlations

The GPPA provides models for all of these quantities.

# 2 INTRODUCTION

## 2.1 The Dielectric Function

Coulomb forces are, by nature, long range and are most appropriately treated in the Fourier transform spaces of wavenumber $\mathbf{q}$ and frequency $\omega$. Of fundamental importance to the treatment of homogeneous Coulomb systems in local thermodynamic equilibrium is the dielectric function $\varepsilon(\mathbf{q},\omega)$, whose definitive properties are expressed by [1],[3],[8]

$$\frac{1}{\varepsilon(\mathbf{q},\omega)} = 1 + u_0(\mathbf{q})\mathrm{K}_{\mathbf{q}}(\omega) = \frac{1}{1 - u_0(\mathbf{q})\Pi_{\mathbf{q}}(\omega)} = \frac{u(\mathbf{q},\omega)}{u_0(\mathbf{q})} \tag{1}$$

where $u_0(\mathbf{q}) = 4\pi e^2/4\pi\varepsilon_0 q^2$ is the bare Coulomb interaction and $u(\mathbf{q},\omega)$ is the corresponding dynamically screened Coulomb potential, which is the effective interaction that incorporates the modifying effects due to polarization of the material medium. The functions $\mathrm{K}_{\mathbf{q}}(\omega)$ and $\Pi_{\mathbf{q}}(\omega)$ are Green functions being respectively the charge-density response function [9],[10] which gives the charge-density response to an applied electric field, and the polarization function [11] which gives the response to the total internal (effective) field. These functions are related by the Dyson equation,

$$\mathrm{K}_{\mathbf{q}}(\omega) = \Pi_{\mathbf{q}}(\omega) + \Pi_{\mathbf{q}}(\omega)u_0(\mathbf{q})\mathrm{K}_{\mathbf{q}}(\omega) \tag{2}$$

which is the screening equation and which yields the screening expansion by iteration. Of these, $\mathrm{K}_{\mathbf{q}}(\omega)$ is the more fundamental. Because $\mathrm{K}_{\mathbf{q}}(\omega)$ describes the charge-density response to an applied external field, it is clearly a retarded Green function, ie it satisfies causality. According to the standard theory of Green functions [12], [13], [14], [11], [1], [2], [15], [16] it therefore possesses a spectral representation in the form,

$$\mathrm{K}_{\mathbf{q}}(\omega) = \frac{1}{2\pi}\int_{-\infty}^{\infty} \frac{\sigma_{\mathbf{q}}(\omega')}{\omega - \omega' + \mathrm{i}0^+}\mathrm{d}\omega' \tag{3}$$

where $\sigma_{\mathbf{q}}(\omega)$ is the spectral density, which is defined on $\mathrm{Im}\,\omega = 0$ to be real and to satisfy

$$\sigma_{\mathbf{q}}(-\omega) = -\sigma_{\mathbf{q}}(\omega) \tag{4}$$

$$\lim_{\omega\to\pm\infty}\left(\omega^2\sigma_{\mathbf{q}}(\omega)\right) = 0 \tag{5}$$

A key property of the spectral density is the f-sum rule

$$\frac{1}{2\pi n}\int_0^{\infty}\sigma_{\mathbf{q}}(\omega)\,\omega\,\mathrm{d}\omega=\frac{q^2}{2m} \tag{6}$$

where $n$ is the particle density. Equation (3) defines $\mathrm{K}_{\mathbf{q}}(\omega)$, and hence, $1/\varepsilon(\mathbf{q},\omega)$, to be an analytic function of a complex variable $\omega$ that is regular (free of singularities) throughout $\mathrm{Im}\,\omega\geq 0$. A contour integral of $\mathrm{K}_{\mathbf{q}}(\omega)$ that encloses the upper half of the complex-$\omega$ plane therefore vanishes, and so, using the property (5), one can deduce the important dispersion relation,

$$\int_{-\infty}^{\infty}\frac{\mathrm{K}_{\mathbf{q}}(\omega')}{\omega'-\omega+\mathrm{i}0^+}\mathrm{d}\omega'=0 \tag{7}$$

Taking the real and imaginary parts of (7) and using the Cauchy identity,

$$\frac{1}{z\pm\mathrm{i}0^+}=\mp\pi\mathrm{i}\,\delta(z)+\wp\left(\frac{1}{z}\right) \tag{8}$$

where $\wp$ denotes principal value, leads to the Kramers-Krönig dispersion relations [17],[18] for $\mathrm{K}_{\mathbf{q}}(\omega)$ on $\mathrm{Im}\,\omega=0$:

$$\begin{aligned}\mathrm{Im}\,\mathrm{K}_{\mathbf{q}}(\omega)&=-\frac{1}{\pi}\wp\int_{-\infty}^{\infty}\frac{\mathrm{Re}\,\mathrm{K}_{\mathbf{q}}(\omega')}{\omega'-\omega}\mathrm{d}\omega'\\ \mathrm{Re}\,\mathrm{K}_{\mathbf{q}}(\omega)&=\frac{1}{\pi}\wp\int_{-\infty}^{\infty}\frac{\mathrm{Im}\,\mathrm{K}_{\mathbf{q}}(\omega')}{\omega'-\omega}\mathrm{d}\omega'\end{aligned} \tag{9}$$

which, upon substituting for $1/\varepsilon(\mathbf{q},\omega)$ from (1), yield the Kramers-Krönig relations for the inverse dielectric function. These equations are also expressed by the Plemelj formula [16]

$$\mathrm{K}_{\mathbf{q}}(\omega)=\frac{1}{2\pi}\wp\int_{-\infty}^{+\infty}\frac{\sigma_{\mathbf{q}}(\omega')}{\omega-\omega'}\mathrm{d}\omega'-\tfrac{1}{2}i\sigma_{\mathbf{q}}(\omega) \tag{10}$$

which follows by direct application of (8) to (3). Making use of (4), equation (10) can be written in the form

$$\mathrm{K}_{\mathbf{q}}(\omega)=\frac{1}{2\pi}\wp\int_{-\infty}^{+\infty}\frac{\omega'\sigma_{\mathbf{q}}(\omega')}{\omega^2-\omega'^2}\mathrm{d}\omega'-\tfrac{1}{2}i\sigma_{\mathbf{q}}(\omega) \tag{11}$$

For large $|\omega|$ we have

$$\mathrm{K}_{\mathbf{q}}(\omega)\sim\frac{1}{2\pi\omega^2}\int_{-\infty}^{+\infty}\omega'\sigma_{\mathbf{q}}(\omega')\mathrm{d}\omega'-\tfrac{1}{2}\mathrm{i}\,\sigma_{\mathbf{q}}(\omega)+\mathcal{O}\left(\omega^{-4}\right) \tag{12}$$

so that, using (5), and by application of the f-sum rule (6),

$$\mathrm{K}_{\mathbf{q}}(\omega) \sim \frac{q^2 n}{m\omega^2} \tag{13}$$

Referring to (1), the high-frequency limit of the inverse dielectric function is deduced to be

$$\frac{1}{\varepsilon(\mathbf{q},\omega)} \sim 1 + \frac{\Omega^2}{\omega^2} \tag{14}$$

where

$$\Omega^2 = \frac{n}{m} q^2 u_0(\mathbf{q}) = \frac{4\pi n}{m} \frac{e^2}{4\pi\varepsilon_0} \tag{15}$$

is the plasma frequency. Equation (14) extends, by analytic continuation, into $\pi \leq \arg\omega \leq 0$. Moreover, the asymptotic form of $\varepsilon(\mathbf{q},\omega)$ is yielded, from (14), as

$$\varepsilon(\mathbf{q},\omega) \sim 1 - \frac{\Omega^2}{\omega^2} \quad : \quad \pi \leq \arg\omega \leq 0 \tag{16}$$

However, the existence of a retarded spectral representation, corresponding to (3), for the polarization function, $\Pi_{\mathbf{q}}(\omega)$, is less obvious, since $\Pi_{\mathbf{q}}(\omega)$ describes the response to the effective (internal) field, rather than to the applied field, so the causality principle is not directly applicable. This problem is extensively discussed in the literature [19], [20], [21], [22]. It turns out that $\Pi_{\mathbf{q}}(\omega)$, and hence $\varepsilon(\mathbf{q},\omega)$, are indeed analytic in the upper half plane, and satisfy Kramers-Krönig dispersion relations similar to $\mathrm{K}_{\mathbf{q}}(\omega)$ and $1/\varepsilon(\mathbf{q},\omega)$, provided that $\varepsilon(\mathbf{q},0) > 1$, which is a condition for stability, as per (30), and so is a property of any physically reasonable dielectric function.

The dielectric function $\varepsilon(\mathbf{q},\omega)$ is therefore a complex function that is regular and free of both singularities and zeros in $\operatorname{Im}\omega \geq 0$. It yields the linear response of the system to an external time-dependent perturbing field $U_0(\mathbf{q},\omega)$ as follows

$$U(\mathbf{q},\omega) = \frac{U_0(\mathbf{q},\omega)}{\varepsilon(\mathbf{q},\omega)} \tag{17}$$

The zeros of $\varepsilon(\mathbf{q},\omega)$ therefore correspond to resonances, which are the collective modes of the system. These are the plasma modes, whose frequencies $\Omega_{\mathbf{q}}$ are defined by the dispersion relation

$$\varepsilon(\mathbf{q},\Omega_{\mathbf{q}}) = 0 \tag{18}$$

The following properties, which follow from the spectral representation above, will be made use of

$$\varepsilon(-\mathbf{q},\omega)=\varepsilon(\mathbf{q},\omega) \tag{19}$$

$$\varepsilon^*(\mathbf{q},\omega)=\varepsilon(\mathbf{q},-\omega^*) \tag{20}$$

$$\lim_{|\omega|\to\infty}\varepsilon(\mathbf{q},\omega)=1-\frac{\Omega^2}{\omega^2}\quad : \ \pi\le\arg\omega\le 0 \tag{21}$$

where $\Omega$ is the plasma frequency, (15). Equation (18) may possess multi-branched solutions. In the case of a simple system with a single mode branch, the plasma frequency is unique and given by, $\Omega=\Omega_{\mathbf{0}}\equiv\lim_{\mathbf{q}\to\mathbf{0}}\Omega_{\mathbf{q}}$.

The dielectric function embodies a full description of the equilibrium properties of a Coulomb system and of the first order (linear) excursions from equilibrium. Complete knowledge of this function enables the equilibrium properties of the material to be calculated. These include

**Equation of state**. The correlation correction to the Hartree-Fock pressure [15], [23], [16] is:

$$\Delta P=-\frac{1}{(2\pi)^4}\int\mathrm{d}^3\mathbf{q}\int_{-\infty}^{+\infty}\mathrm{d}\,\omega\frac{1}{\mathrm{e}^{\omega/T}-1}\mathrm{Im}\left(\frac{1}{\varepsilon_\lambda(\mathbf{q},\omega)}+\varepsilon_\lambda(\mathbf{q},\omega)\right)\int_0^1\frac{\mathrm{d}\lambda}{\lambda} \tag{22}$$

where $\varepsilon_\lambda(\mathbf{q},\omega)=\varepsilon(\mathbf{q},\omega)\big|e^2\to\lambda e^2$

**Refractive index** (in long-wavelength limit):

$$n(\omega)^2=\tfrac{1}{2}\left(\mathrm{Re}\left(\varepsilon(\mathbf{0},\omega)\right)+\left|\varepsilon(\mathbf{0},\omega)\right|\right) \tag{23}$$

**Electrical conductivity** (in long-wavelength limit):

$$\sigma(\omega)=\mathrm{i}\,\omega\varepsilon_0\left(1-\varepsilon(\mathbf{0},\omega)\right) \tag{24}$$

**Absorption coefficient** (in long-wavelength limit):

$$\rho\kappa(\omega)=\frac{\omega}{n(\omega)c}\mathrm{Im}\left(\varepsilon(\mathbf{0},\omega)\right) \tag{25}$$

**Charge-density dynamic structure factor** [3]

$$\begin{aligned}S(\mathbf{q},\omega)&=\frac{1}{\pi n_{\mathrm{e}}u_0(q)}\frac{1}{1-\mathrm{e}^{-\omega/T}}\mathrm{Im}\left(\frac{-1}{\varepsilon(\mathbf{q},\omega)}\right)\\&=\frac{q^2}{\pi m_{\mathrm{e}}\Omega_{\mathrm{e}}{}^2}\frac{1}{1-\mathrm{e}^{-\omega/T}}\mathrm{Im}\left(\frac{-1}{\varepsilon(\mathbf{q},\omega)}\right)\end{aligned} \tag{26}$$

which is a statement of the Fluctuation Dissipation Theorem [3], [24], [8] where, in order to allow generalization to multicomponent plasmas, we have chosen to adopt the convention whereby the structure factor is normalized to the electron density, $n_e$; and where $u_0(q) = 4\pi e^2/4\pi\varepsilon_0 q^2$ denotes the Coulomb potential between two unit charges.

**Charged particle stopping power** [16], [25], [26]

$$\frac{\mathrm{d}}{\mathrm{d}t}\langle E_\mathrm{p}\rangle = -\frac{2}{\pi v}\left(\frac{Z^2e^2}{4\pi\varepsilon_0}\right)\int_0^\infty \mathrm{Im}\left(\frac{-1}{\varepsilon(\mathbf{q},\omega)}\right)\omega\,\mathrm{d}\omega\int_{q_-}^{q_+}\frac{\mathrm{d}q}{q} \tag{27}$$

where $m$, $Ze$, $E_\mathrm{p}$, $v$ are respectively the reduced mass (for collisions with electrons) charge, kinetic energy and velocity of the incident particles and where $q_\pm = \sqrt{m^2v^2+2m\omega}\pm mv$. For fast particles, the integral is dominated by small-$q$ scatterings, and so

$$\begin{aligned}\frac{\mathrm{d}}{\mathrm{d}t}\langle E\rangle &\simeq -\frac{2}{\pi v}\left(\frac{Z^2e^2}{4\pi\varepsilon_0}\right)\int_0^\infty \mathrm{Im}\left(\frac{-1}{\varepsilon(\mathbf{0},\omega)}\right)\ln\left(\frac{q_+}{q_-}\right)\omega\,\mathrm{d}\omega \\ &\simeq -\frac{2}{\pi v}\left(\frac{Z^2e^2}{4\pi\varepsilon_0}\right)\int_0^\infty \mathrm{Im}\left(\frac{-1}{\varepsilon(\mathbf{0},\omega)}\right)\ln\left(\frac{2mv^2}{\omega}\right)\omega\,\mathrm{d}\omega\end{aligned} \tag{28}$$

which is related to the Bethe fast-ion stopping power [16], [27], [28], [29], [30]

**Isothermal compressibility**

The polarization function gives the change in density due to a change in the total (effective) field, which, in static situations, is equivalent to a change in the bulk pressure. The isothermal compressibility $\mathfrak{K}$ of a one-component plasma consisting of particles of charge $e$ is therefore given by

$$\Pi_\mathbf{0}(0) = -\frac{\partial n}{\partial \mu} = -n\frac{\partial n}{\partial P} == -n^2\mathfrak{K} \tag{29}$$

In terms of the dielectric function, this yields

$$\begin{aligned}\mathfrak{K} &= \frac{1}{n^2}\lim_{\mathbf{q}\to\mathbf{0}}\frac{\varepsilon(\mathbf{q},0)-1}{u_0(\mathbf{q})} \\ &= \frac{1}{n\,m\,\Omega^2}\lim_{\mathbf{q}\to\mathbf{0}}\left(q^2\left(\varepsilon(\mathbf{q},0)-1\right)\right)\end{aligned} \tag{30}$$

## 2.2 Sum Rules

In accordance with equations (245) and (246) in APPENDIX A, the above properties imply the following sum rules:

- Conductivity sum rule:

$$\int_{-\infty}^{\infty} \operatorname{Im}\varepsilon(\mathbf{q},\omega)\,\omega\mathrm{d}\omega = \pi\Omega^2 \tag{31}$$

- Compressibility sum-rule:

$$\int_{-\infty}^{\infty} \operatorname{Im}\varepsilon(\mathbf{q},\omega)\frac{\mathrm{d}\omega}{\omega} = \pi\alpha_{\mathbf{q}} \tag{32}$$

- Generalized f-sum rule:

$$\int_{-\infty}^{\infty} \operatorname{Im}\left(\frac{-1}{\varepsilon(\mathbf{q},\omega)+\beta}\right)\omega\mathrm{d}\omega = \pi\frac{\Omega^2}{(1+\beta)^2} \tag{33}$$

- Generalized screening sum rule:

$$\int_{-\infty}^{\infty} \operatorname{Im}\left(\frac{-1}{\varepsilon(\mathbf{q},\omega)+\beta}\right)\frac{\mathrm{d}\omega}{\omega} = \pi\frac{\alpha_{\mathbf{q}}}{(1+\alpha_{\mathbf{q}}+\beta)(1+\beta)} \tag{34}$$

In the above, $\beta$ is a real positive semidefinite ($\beta \geq 0$) parameter, independent of $\omega$; and

$$\alpha_{\mathbf{q}} = \varepsilon(\mathbf{q},0) - 1 \tag{35}$$

Since $\operatorname{Im}\varepsilon(\mathbf{q},-\omega) = -\operatorname{Im}\varepsilon(\mathbf{q},\omega)$ on $\operatorname{Im}\omega = 0$, the above also yield the following sum rules as integrals from 0 to $\infty$:

- Conductivity sum rule:

$$\int_{0}^{\infty} \operatorname{Im}\varepsilon(\mathbf{q},\omega)\,\omega\mathrm{d}\omega = \frac{\pi}{2}\Omega^2 \tag{36}$$

- Compressibility sum-rule:

$$\int_{0}^{\infty} \operatorname{Im}\varepsilon(\mathbf{q},\omega)\frac{\mathrm{d}\omega}{\omega} = \frac{\pi}{2}\alpha_{\mathbf{q}} \tag{37}$$

- F-sum rule:

$$\int_0^\infty \mathrm{Im}\left(\frac{-1}{\varepsilon(\mathbf{q},\omega)}\right)\omega\,\mathrm{d}\omega = \frac{\pi}{2}\Omega^2 \tag{38}$$

- Screening sum rule:

$$\int_0^\infty \mathrm{Im}\left(\frac{-1}{\varepsilon(\mathbf{q},\omega)}\right)\frac{\mathrm{d}\omega}{\omega} = \frac{\pi}{2}\frac{\alpha_{\mathbf{q}}}{\left(1+\alpha_{\mathbf{q}}\right)} \tag{39}$$

Note that the underlying origin of the f-sum rule is equation (6), from which the asymptotic form of the dielectric function is derived, rather than any mathematical proof in APPENDIX A. This is because, in the above formulation, the asymptotic form of the dielectric function, upon which any mathematical proof would depend, is derived from the properties of the spectral density, in particular the f-sum rule, which leads to equation (13). A “proof” of the f-sum rule according to APPENDIX A would therefore constitute a circular argument. However, having established the asymptotic form of the dielectric function, the other sum rules follow from the arguments in APPENDIX A.

## 2.3 Static Properties

The static properties of the system are summed up by the static dielectric function $\varepsilon(\mathbf{q},0)=1+\alpha_{\mathbf{q}}$ and the static structure factor

$$S(\mathbf{q}) = \int_{-\infty}^{\infty} S(\mathbf{q},\omega)\,\mathrm{d}\omega \tag{40}$$

Introducing the *classical* Debye length, $D$, defined by

$$D^2 = \frac{T}{m\Omega^2} \tag{41}$$

and defining the functions $\tilde{R}_{\mathbf{q}}$, $R_{\mathbf{q}}$ and $S_{\mathbf{q}}$ by,

$$\tilde{R}_{\mathbf{q}} = \frac{2}{\pi}q^2D^2\int_0^\infty \mathrm{Im}\,\varepsilon(\mathbf{q},\omega)\frac{\mathrm{d}\omega}{\omega} \tag{42}$$

$$R_{\mathbf{q}} = \frac{2}{\pi}q^2D^2\int_0^\infty \frac{\omega}{2T}\coth\left(\frac{\omega}{2T}\right)\mathrm{Im}\,\varepsilon(\mathbf{q},\omega)\frac{\mathrm{d}\omega}{\omega} \tag{43}$$

$$S_{\mathbf{q}} = \frac{2}{\pi}q^2D^2\int_0^\infty \frac{\omega}{2T}\coth\left(\frac{\omega}{2T}\right)\mathrm{Im}\left(\frac{-1}{\varepsilon(\mathbf{q},\omega)}\right)\frac{\mathrm{d}\omega}{\omega} \tag{44}$$

then, according to the compressibility sum-rule (37),

$$\tilde{R}_{\mathbf{q}} = q^2 D^2 \alpha_{\mathbf{q}} \equiv q^2 D^2 \left( \varepsilon(\mathbf{q},0) - 1 \right) \tag{45}$$

and hence

$$\varepsilon(\mathbf{q},0) = 1 + \frac{\tilde{R}_{\mathbf{q}}}{q^2 D^2}$$

$$\Pi_{\mathbf{q}}(0) = -\frac{n}{T}\tilde{R}_{\mathbf{q}} \tag{46}$$

which are exact formulae for the static dielectric and polarization functions. In section 3.1, these will be shown to reduce to their Debye-Hückel forms, for small wavenumbers only, in weakly-coupled plasmas.

A connection between $\tilde{R}_{\mathbf{q}}$ and $R_{\mathbf{q}}$ is provided in the high-temperature/ small-$q$ limit by expanding $\frac{\omega}{2T}\coth\left(\frac{\omega}{2T}\right)$ in the integrand of (63) about $\omega = 0$. This yields

$$R_{\mathbf{q}} = \frac{2}{\pi} q^2 D^2 \int_0^\infty \left( 1 + \frac{1}{3}\left(\frac{\omega}{2T}\right)^2 + \ldots \right) \operatorname{Im}\varepsilon(\mathbf{q},\omega) \frac{\mathrm{d}\omega}{\omega}$$

$$= \tilde{R}_{\mathbf{q}} + \frac{1}{6\pi}\frac{q^2 D^2}{T^2} \int_0^\infty \omega \operatorname{Im}\varepsilon(\mathbf{q},\omega)\, \mathrm{d}\omega + \ldots \tag{47}$$

$$= \tilde{R}_{\mathbf{q}} + \frac{q^2}{12mT} + \ldots$$

and hence $R_{\mathbf{q}}$ and $\tilde{R}_{\mathbf{q}}$ tend to a common limit,

$$\lim_{\mathbf{q}\to\mathbf{0}} \tilde{R}_{\mathbf{q}} = \lim_{\mathbf{q}\to\mathbf{0}} R_{\mathbf{q}} = R_{\mathbf{0}} \tag{48}$$

Equation (46) now yields the isothermal compressibility (30) as follows

$$\mathfrak{K} = \lim_{q\to 0} \frac{\alpha_{\mathbf{q}}}{n^2 u_0(q)} = \frac{R_{\mathbf{0}}}{nT} \tag{49}$$

which again is an exact result.

For large *q,* we find that

$$\lim_{q\to\infty} \tilde{R}_{\mathbf{q}} = 0, \quad \lim_{q\to\infty} R_{\mathbf{q}} = 1 \tag{50}$$

For the static structure factor, we have, using equation (26) and the fact, which follows from (20), that $\mathrm{Im}\,\varepsilon(\mathbf{q},\omega)$ is an odd function of $\omega$ on $\mathrm{Im}\,\omega = 0$,

$$S(\mathbf{q}) = \int_0^\infty \left(1 + \mathrm{e}^{-\omega/T}\right) S(\mathbf{q},\omega)\,\mathrm{d}\omega$$

$$= \frac{n}{n_\mathrm{e}} \frac{2}{\pi} q^2 D^2 \int_0^\infty \frac{\omega}{2T} \coth\left(\frac{\omega}{2T}\right) \mathrm{Im}\left(\frac{-1}{\varepsilon(\mathbf{q},\omega)}\right) \frac{\mathrm{d}\omega}{\omega} \tag{51}$$

$$= \frac{n}{n_\mathrm{e}} S_\mathbf{q}$$

In a similar fashion,

$$\int_{-\infty}^\infty S(\mathbf{q},\omega)\,\omega\,\mathrm{d}\omega = \int_0^\infty \left(1 - \mathrm{e}^{-\omega/T}\right) S(\mathbf{q},\omega)\,\omega\,\mathrm{d}\omega$$

$$= \frac{n}{n_\mathrm{e}} \frac{q^2}{\pi m \Omega^2} \int_0^\infty \mathrm{Im}\left(\frac{-1}{\varepsilon(\mathbf{q},\omega)}\right) \omega\,\mathrm{d}\omega \tag{52}$$

$$= \frac{n}{n_\mathrm{e}} \frac{q^2}{2m}$$

which is a straightforward application of the f-sum rule (38).

The functions $R_\mathbf{q}$ and $\tilde{R}_\mathbf{q}$ play an important role in the general theory. It has already been demonstrated that $\tilde{R}_\mathbf{q}$ provides the short-wavelength corrections to the dielectric and polarization functions according to the (exact ) formulae (46), while, in the next section, it is found that, in the context of the Random Phase Approximation, the function $R_\mathbf{q}$ yields the exact static structure factor for a system of non-interacting indistinguishable (Fermi) particles.

## 3 APPROXIMATE REPRESENTATIONS OF THE DIELECTRIC FUNCTION

### 3.1 Random Phase Approximation

A fully self-consistent approximation that applies to dilute (weakly coupled) systems interacting via long-range forces, such that any short-range correlations induced by the interactions are negligible, is the Random Phase Approximation (RPA) [1], [2], [3], [16], [31] also sometimes known as the Ring Approximation. The polarization part $\Pi_\mathbf{q}(\omega)$ of the Coulomb Green function, is given, in the RPA, by the well-known Lindhard formula [32]

$$\Pi_{\mathbf{q}}^{0}=\frac{1}{\mathfrak{V}}\sum_{\mathbf{k}}\frac{p(\mathbf{k})-p(\mathbf{k}-\mathbf{q})}{E(\mathbf{k})-E(\mathbf{k}-\mathbf{q})-\omega-\mathrm{i}0^{+}} \tag{53}$$

where $E(\mathbf{k})$ is the energy of the single-particle fermion state corresponding to wavevector $\mathbf{k}$, and $p(\mathbf{k})$ is the probability, given by the Fermi-Dirac distribution,

$$p(\mathbf{k})=\frac{1}{1+\exp\left(E(\mathbf{k})/T-\eta\right)} \tag{54}$$

of the state being occupied. The RPA describes systems of point-like Coulomb particles in the weak-coupling limit, and provides an exact description of quantum statistical effects (degeneracy and exchange correlations) in this limit. The RPA therefore provides an important baseline approximation for Coulomb systems.

A closed form expression of the imaginary part of the RPA dielectric function, $\varepsilon^{0}$, for a system comprising a single fermion species of mass $m$, is given in terms of the dimensionless parameters, $u=\omega/T$ and $v=q^{2}/2mT$, as follows [33], [16]

$$\operatorname{Im}\varepsilon^{0}(\mathbf{q},\omega)\equiv\operatorname{Im}\varepsilon^{0}(u,v)=\frac{\pi}{8}\frac{\Omega^{2}}{T^{2}}\frac{1}{v^{3/2}}\frac{1}{I_{1/2}(\eta)}L(u,v;\eta) \tag{55}$$

where $\Omega^{2}=ne^{2}/\varepsilon_{0}m$ is the plasma frequency, $\eta=\mu/T$ is the degeneracy parameter, $I_{1/2}(\eta)$ is the standard Fermi integral as defined by

$$I_{j}(x)=\int_{0}^{\infty}\frac{y^{j}}{1+\exp(y-x)}\mathrm{d}y \tag{56}$$

and

$$L(u,v;\eta)=\ln\left(\frac{1+\exp\left(\eta-(u-v)^{2}/4v\right)}{1+\exp\left(\eta-(u+v)^{2}/4v\right)}\right) \tag{57}$$

A useful universal approximation, which applies in most regimes, ie generally, except that of extreme degeneracy where $\eta\gg1$, is

$$L(u,v;\eta)\simeq\frac{2\sinh\left(\frac{1}{2}u\right)}{1+\exp\left(u^{2}/4v+v/4-\eta\right)} \tag{58}$$

Equation (58) incorporates the same limiting forms as (57) for $v\to0$ $(u\neq0)$, $v\to\infty$, $u\to0$, $|u|\to\infty$ $(u^{2}\gg4\eta v)$, and $\eta\to-\infty$. It also satisfies the inequality (for $u>0$):

$$\frac{2\sinh\left(\frac{1}{2}u\right)}{\exp\left(\frac{1}{2}u\right)+\exp\left(u^2/4v+v/4-\eta\right)} < L(u,v;\eta) < \frac{2\sinh\left(\frac{1}{2}u\right)}{\exp\left(-\frac{1}{2}u\right)+\exp\left(u^2/4v+v/4-\eta\right)} \tag{59}$$

which is an exact property of (57). The low temperature, extreme degeneracy limit of (57) is determined by writing $\tilde{L}(r,s;\eta)=\eta^{-1}L(u,v;\eta)$ where $r=u/2\eta$, $s^2=v/\eta$ and taking the limit $\eta\to\infty$ , which yields the more familiar low-temperature forms of the formula, eg. [15], namely, $\tilde{L}\to 2r$ , $\tilde{L}\to 1-(r/s-s/2)^2$, $\tilde{L}\to 0$ in the respective domains $(r/s+s/2)^2<1$ , $(r/s-s/2)^2<1<(r/s+s/2)^2$ , $(r/s-s/2)^2>1$ where $r=\omega/2T_\mathrm{F}$, $s=q/k_\mathrm{F}$ .

An important property of the general function $L(u,v;\eta)$ is that it is independent of the interactions between the particles. The function $L(u,v;\eta)$ therefore relates to the properties of the finite-temperature ideal Fermi gas, which is a system comprising non-interacting fermions in which the only correlations are due to exchange. This is most evident when one considers the dynamic structure factor (26), given by the dielectric superposition principle [34],

$$S(\mathbf{q},\omega)=\frac{S^0(\mathbf{q},\omega)}{\left|\varepsilon(\mathbf{q},\omega)\right|^2} \tag{60}$$

where, in the RPA,

$$\begin{aligned} S^0(\mathbf{q},\omega) &= \frac{q^2}{\pi m_\mathrm{e}\Omega_\mathrm{e}^{\,2}}\frac{1}{1-\mathrm{e}^{-\omega/T}}\,\mathrm{Im}\,\varepsilon^0(\mathbf{q},\omega) \\ &= \frac{n}{n_\mathrm{e}}\frac{1}{4Tv^{1/2}}\frac{1}{I_{1/2}(\eta)}\frac{1}{1-\mathrm{e}^{-u}}L(u,v;\eta) \end{aligned} \tag{61}$$

which, since it is manifestly independent of the particle interactions, is the dynamic structure factor for an ideal (non-interacting) Fermi gas. Combining equations (60) and (61), with $\varepsilon(\mathbf{q},\omega)=\varepsilon^0(\mathbf{q},\omega)$, yields the fluctuation dissipation theorem (26) in terms of the RPA structure factor and dielectric function. However equation (60) can hold in equilibrium situations beyond RPA provided that $\mathrm{Im}\,\varepsilon(\mathbf{q},\omega)$ can be taken to be proportional to the square of the particle's electric charge, and moreover is, unlike (61), not confined to equilibrium situations. The derivation of (60) given by Ichimaru [34], for example, does not explicitly depend upon the RPA as providing the underlying description.

The static structure factor of the ideal Fermi gas that follows from (61) is

$$S^0(\mathbf{q}) \equiv \int_{-\infty}^{+\infty} S^0(\mathbf{q},\omega)\,\mathrm{d}\omega = \frac{n}{n_\mathrm{e}}\frac{1}{4v^{1/2}}\frac{1}{I_{1/2}(\eta)}\int_{-\infty}^{+\infty}\frac{1}{1-\mathrm{e}^{-u}}L(u,v;\eta)\,\mathrm{d}u$$

$$= \frac{n}{n_\mathrm{e}}\frac{1}{4v^{1/2}}\frac{1}{I_{1/2}(\eta)}\int_0^\infty L(u,v;\eta)\coth\left(\tfrac{1}{2}u\right)\mathrm{d}u \tag{62}$$

$$= \frac{n}{n_\mathrm{e}}R_\mathbf{q}$$

where, in accordance with (43) and (55),

$$R_\mathbf{q} = \frac{1}{2v^{1/2}}\frac{1}{I_{1/2}(\eta)}\int_0^\infty L(u,v;\eta)\tfrac{1}{2}u\coth\left(\tfrac{1}{2}u\right)\frac{\mathrm{d}u}{u} \tag{63}$$

where $D$ is the classical Debye length, defined by (41).

The integral in (63) can be evaluated exactly, in the limit $q \to 0$, to yield,

$$R_\mathbf{0} = \lim_{v\to 0}\frac{1}{2v^{1/2}}\frac{1}{I_{1/2}(\eta)}\int_0^\infty L(u,v;\eta)\tfrac{1}{2}u\coth\left(\tfrac{1}{2}u\right)\frac{\mathrm{d}u}{u} = \frac{I'_{1/2}(\eta)}{I_{1/2}(\eta)} = \tfrac{1}{2}T\left\langle E^{-1}\right\rangle \tag{64}$$

where $\langle\ \rangle$ denotes the thermal average, taken over a Fermi-Dirac distribution of free particles with energies given by $E(\mathbf{k}) = k^2/2m$.

In the Boltzmann limit, $\eta \ll 0$, $1-R_\mathbf{q}$ is exponentially small ($\ln\left(1-R_\mathbf{q}\right) = \mathcal{O}(\eta)$) while, in the limit of extreme degeneracy, $\eta \to \infty$,

$$\begin{aligned} R_\mathbf{q} &= \tfrac{3}{4}s\left(1-\tfrac{1}{12}s^2\right) &&: \quad 0 \le s \le 2 \\ &= 1 &&: \quad s > 2 \end{aligned} \tag{65}$$

where

$$s^2 = \tfrac{2}{3}vR_\mathbf{0} = \frac{q^2}{6m}\left\langle E^{-1}\right\rangle \tag{66}$$

However one cannot extrapolate from (65) to the finite-$\eta$ regime because of the singular behaviour of the function in the limit; particularly because of the non-commutation of the limits $\eta \to \infty$ and $s \to 0$.

Further integral relations satisfied by $L(u,v;\eta)$ follow from the conductivity sum-rule (36) and the compressibility sum-rule (37). The former implies

$$\int_0^\infty L(u,v;\eta)\,u\,\mathrm{d}u = 4v^{3/2} I_{1/2}(\eta) \tag{67}$$

together with

$$\int_{-\infty}^\infty S^0(\mathbf{q},\omega)\,\omega\,\mathrm{d}\omega = \frac{n}{n_\mathrm{e}}\frac{q^2}{2m} \tag{68}$$

which is just a special case of (52). Now consider the function $\tilde{R}_\mathbf{q}$ defined by (42), which, when combined with (55), yields

$$\tilde{R}_\mathbf{q} = \frac{1}{2v^{1/2}}\frac{1}{I_{1/2}(\eta)}\int_0^\infty L(u,v;\eta)\frac{\mathrm{d}u}{u} \tag{69}$$

As is shown in Appendix B.1, for a Boltzmann plasma,

$$\tilde{R}_\mathbf{q} = \frac{2}{\sqrt{v}}\Phi\left(\frac{\sqrt{v}}{2}\right) \tag{70}$$

where $\Phi(x)$ is Dawson's Integral,

$$\Phi(x) = \mathrm{e}^{-x^2}\int_0^x \mathrm{e}^{t^2}\,\mathrm{d}t \tag{71}$$

while, in the limit of extreme degeneracy, $\eta >> 1$,

$$\begin{aligned}\tilde{R}_\mathbf{q} &= R_\mathbf{0}\frac{\eta^{1/2}}{v^{1/2}}\left[s + \left(1 - \frac{s^2}{4}\right)\ln\left(\left|\frac{s+2}{s-2}\right|\right)\right] \\ &= \tfrac{1}{2}R_0\left[1 + \frac{1}{s}\left(1 - \frac{s^2}{4}\right)\ln\left(\left|\frac{s+2}{s-2}\right|\right)\right]\end{aligned} \tag{72}$$

which results, fairly straightforwardly, from performing the integration with the limiting form of the function $L(u,v;\eta\to\infty)$ given above.

A universal Padé approximant for $\tilde{R}_\mathbf{q}$ is

$$\tilde{R}_\mathbf{q} = R_\mathbf{0}\frac{1 + \frac{1}{16}s^2 + \frac{1}{48}t^2}{1 + \frac{1}{16}s^2 + \frac{5}{48}t^2 + \frac{15}{256}s^4 + \frac{1}{256}t^4} \tag{73}$$

where

$$s^2 = \frac{q^2}{6m}\left\langle E^{-1}\right\rangle; \qquad t^2 = \frac{q^2}{2m}\left\langle\left\langle E^{-1}\right\rangle\right\rangle \tag{74}$$

where $\langle\langle f \rangle\rangle \equiv \langle f q \rangle / \langle q \rangle$, $q = 1 - p$. In the degenerate limit, $s^2 = t^2 = q^2/2mT_{\mathrm{F}} = \left(q^2/3mT\right)R_{\mathbf{0}} = v/\eta$, which then yields,

$$\tilde{R}_{\mathbf{q}} \sim R_{\mathbf{0}} \frac{1+\frac{1}{12}s^2}{1+\frac{1}{6}s^2+\frac{1}{16}s^4} \tag{75}$$

which is recognised as an approximation to (72). In the Boltzmann limit, $s^2 = q^2/3mT = 2v/3$; $t^2 = q^2/mT = 3s^2 = 2v$, which yields

$$\tilde{R}_{\mathbf{q}} \sim R_{\mathbf{0}} \frac{1+\frac{1}{12}v}{1+\frac{1}{4}v+\frac{1}{24}v^2} \tag{76}$$

which represents an approximation to (70). Equations (73), and hence (75) and (76), are constructed to give the correct low-order expansions to $\mathcal{O}\left(s^2\right)$ and the correct asymptotic forms as $s \to \infty$, in both the degenerate and non-degenerate limits.

These results show that the static dielectric function (46), evaluated in the RPA, reduces to the generalized (degeneracy-corrected) Debye-Hückel formula, $\varepsilon(\mathbf{q},0) = 1 + R_{\mathbf{0}}/q^2D^2$, only for small wavenumbers, when $q^2 << mT/R_{\mathbf{0}}$. The generalized Debye-Hückel formula therefore gives only the long range part of the screening potential in weakly coupled systems. The quantum (exchange) corrections at short distances, which are those $\lesssim$ the thermal deBroglie wavelength $\left(R_0/mT\right)^{1/2}$, are expressed by the function $\tilde{R}_{\mathbf{q}}/R_{\mathbf{0}}$. The following figures illustrate this function in the non-degenerate (Boltzmann) and fully degenerate limits.

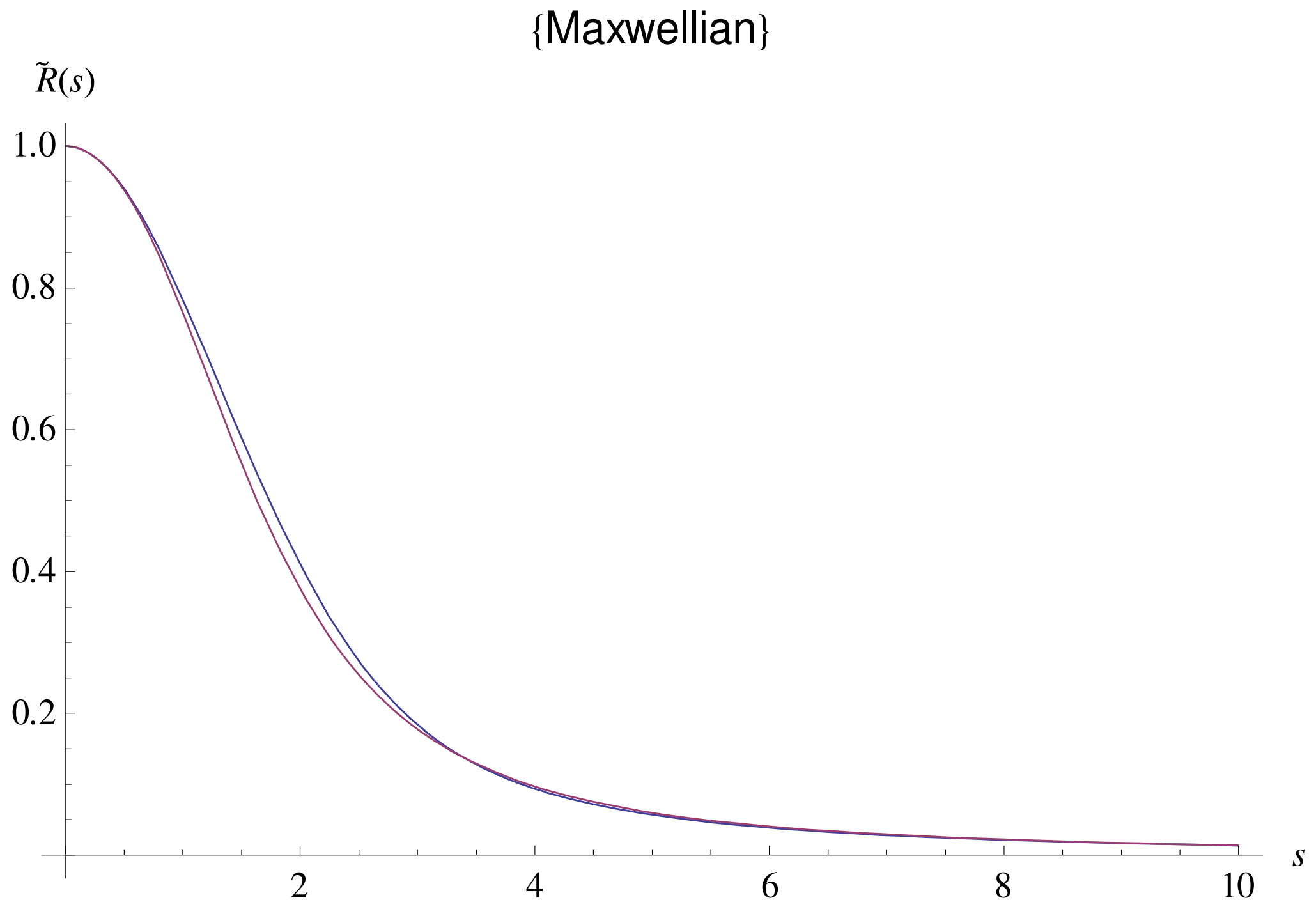

**Figure 1:** Plot of the function $\tilde{R}(s)=\tilde{R}(s)/R_0$, in the Boltzmann limit, as given by equation (70), with $s^2=2v/3$, (blue/continuous curve) compared with Padé approximant (76) (red dashed curve).

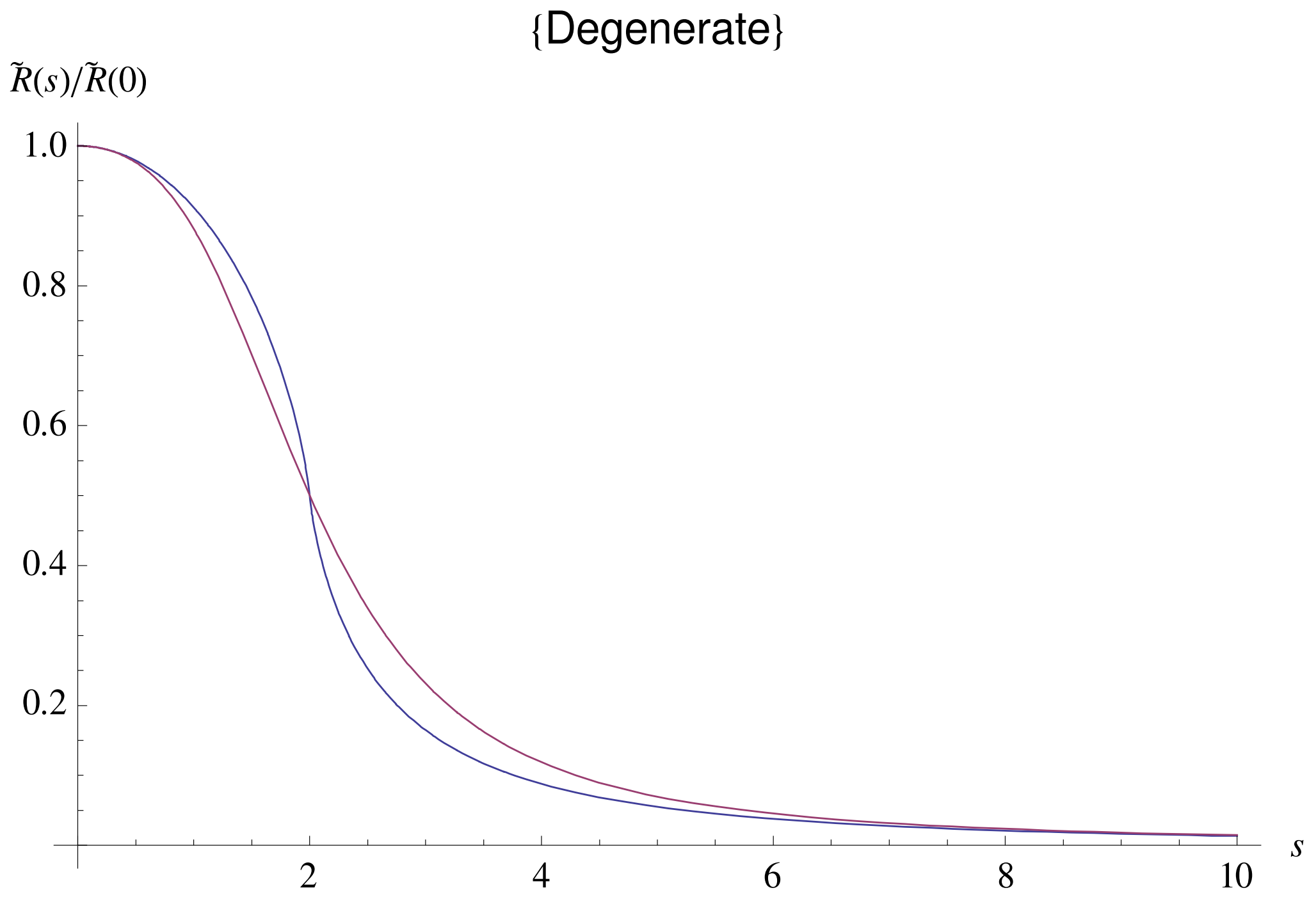

**Figure 2:** Plot of the RPA function $\tilde{R}(s)/R_0$ in the degenerate limit, as given by equation (72), (blue/continuous curve) compared with Padé approximant (75) (red/dashed curve).

The expressions given above for the functions are those given by the RPA for weakly coupled systems. For strongly coupled systems, some aspects of the qualitative behaviour can be expected to be retained. In particular, the rapid fall-off at large $q$ ensures convergence of some physically-significant integrals over $\mathbf{q}$ that, in the Debye-Hückel approximation, for example, would otherwise be divergent.

In the classical limit (small $\mathbf{q}$) the Lindhard formula (53) becomes $\Pi^0_{\mathbf{q}} = \left(\Omega^2/u_0(\mathbf{q})\right)\left\langle\left(\omega\left(\omega+\mathrm{i}0^+\right)-v^2q^2\right)^{-1}\right\rangle$ where $v = k/m$ is the particle velocity, which corresponds, according to (1), to the dielectric function, $\varepsilon^{\mathrm{cl}}(\mathbf{q},\omega) = 1-\Omega^2\left\langle\left(\omega\left(\omega+\mathrm{i}0^+\right)-v^2q^2\right)^{-1}\right\rangle$. In the static limit, this reduces to $\varepsilon^{\mathrm{cl}}(\mathbf{q},0) = 1+\left(\Omega^2/q^2\right)\left\langle v^{-2}\right\rangle \equiv 1+R_0/q^2D^2$, which is the (degeneracy corrected) Debye-Hückel formula, as remarked earlier, and which incorporates the identity $\left\langle v^{-2}\right\rangle = R_0 m/T$, cf. equation (64). At high frequencies, the dielectric function, $\varepsilon^{\mathrm{cl}}(\mathbf{q},\omega)$ becomes

$$\varepsilon^{cl}(\mathbf{q},\omega) \sim 1-\frac{\Omega^2}{\omega^2}\left(1+\frac{q^2}{\omega^2}\langle v^2\rangle\right)$$

$$=1-\frac{\Omega^2}{\omega^2}\left(1+\frac{\Omega^2 D^2 q^2}{\omega^2}\langle v^2\rangle\left\langle\frac{1}{v^2}\right\rangle\right) \quad (77)$$

$$=1-\frac{\Omega^2}{\omega^2}\left(1+\gamma(\eta)\frac{\Omega^2 D^2 q^2}{\omega^2}\right)$$

where

$$\gamma(\eta)=\langle E\rangle\langle E^{-1}\rangle=\frac{2I_{3/2}(\eta)\,I'_{1/2}(\eta)}{\left(I_{1/2}(\eta)\right)^2} \quad (78)$$

which lies in the range $9/5$ (fully degenerate) to $3$ (Boltzmann) [35].

The collective modes $\Omega_{\mathbf{q}}$ of a plasma are given by the zeros of the dielectric function, ie $\varepsilon(\mathbf{q},\Omega_{\mathbf{q}})=0$. In the case of the dielectric function (77), the dispersion relation, for small values of $\mathbf{q}$, is

$$\Omega_{\mathbf{q}}^{\,2} \simeq \Omega^2\left(1+\frac{\gamma(\eta)}{\alpha_{\mathbf{q}}}\right) \quad (79)$$

which is the classical dispersion relation for the collective modes of a collisionless plasma.

The formula

$$\Omega_{\mathbf{q}}^{\,2} \simeq \Omega^2\left(1+\frac{3}{\alpha_{\mathbf{q}}}\right) \quad (80)$$

is a reasonable fit to the zeros of the RPA dielectric function, over the (finite) range of $\mathbf{q}$ for which Landau damping is weak, but not negligible, and where $\Omega_{\mathbf{q}}$ is the modified frequency defined by (90), in both Boltzmann and fully degenerate regimes. Equation (80) is a form of the Bohm-Gross relation [35].

### 3.1.1 Random Phase Approximation Treatment of Multicomponent Plasmas

The above properties of a plasma in the Random Phase Approximation apply to a single charge species. However when the approximation is extended to multi-component plasmas, which are plasmas containing more than one type of ion species $\{a\}$, the RPA polarization function $\Pi^0_{\mathbf{q}}$ appearing in the screening equation (2) is representable as a sum of the polarization parts $\Pi_{\mathbf{q}}^{a}$ of each component, which, in the RPA, are irreducible free-particle-hole 'bubble' diagrams in which only one type of particle can appear. As a result, the RPA dielectric function of a multi-component plasma is just a simple sum over the dielectric functions $\varepsilon_a(\mathbf{q},\omega)$ of each component species, calculated as if the other components were not present, and expressed as follows,

$$\varepsilon(\mathbf{q},\omega) = 1 + \sum_a \left(\varepsilon_a(\mathbf{q},\omega) - 1\right) \tag{81}$$

Some particular consequences of this, which follow from (21), (41) and (69), are

$$\Omega^2 = \sum_a {\Omega_a}^2 \tag{82}$$

$$\frac{\tilde{R}_{\mathbf{q}}}{D^2} = \sum_a \frac{\tilde{R}_{\mathbf{q}}^{a}}{{D_a}^2} \tag{83}$$

$$\alpha_{\mathbf{q}} = \sum_a \alpha_{\mathbf{q}}^{a} = \sum_a \frac{\tilde{R}_{\mathbf{q}}^{a}}{q^2 {D_a}^2} \tag{84}$$

where ${\Omega_a}^2 = \dfrac{4\pi {Z_a}^2 e^2 n_a}{4\pi\varepsilon_0 m_a}$ and ${D_a}^2 = \dfrac{T}{m_a {\Omega_a}^2}$ are respectively the squares of the plasma frequency and the Debye length associated with species $a$.

### 3.2 Plasmon Pole Approximation

Another approximate representation of the dielectric function, one which focuses on the resonances, is the plasmon pole approximation [36],[37], which, in its standard form, represents the dielectric function of a single component as follows

$$\frac{1}{\varepsilon_a(\mathbf{q},\omega)} = 1 + \frac{{\Omega_a}^2}{\left(\omega + \mathrm{i}0^+\right)^2 - \left(\Omega_{\mathbf{q}}^{a}\right)^2} \tag{85}$$

By virtue of the Cauchy identity,

$$\mathrm{Im}\left(\frac{1}{z - \mathrm{i}0^+}\right) = -\mathrm{Im}\left(\frac{1}{z + \mathrm{i}0^+}\right) = \pi\delta(z) \tag{86}$$

this gives

$$\mathrm{Im}\left(\frac{-1}{\varepsilon(\mathbf{q},\omega)}\right) = \frac{\pi}{2}\frac{{\Omega_a}^2}{\Omega_{\mathbf{q}}^{a}}\left[\delta\left(\omega - \Omega_{\mathbf{q}}^{a}\right) - \delta\left(\omega + \Omega_{\mathbf{q}}^{a}\right)\right] \tag{87}$$

Then applying the screening sum-rule, (39), or directly from (85), one finds a direct relationship between the parameters $\Omega_{\mathbf{q}}$ and $\alpha_{\mathbf{q}}$:

$$\left(\Omega_{\mathbf{q}}^{a}\right)^{2}=\Omega_{a}^{\;2}\left(1+\frac{1}{\alpha_{\mathbf{q}}^{a}}\right) \tag{88}$$

Unfortunately for this approximation, the relation (88) is inconsistent with the equivalent RPA formulae given by equations (79) - (80).

Equation (85) represents the collective plasma mode in terms of real frequencies $\Omega_{\mathbf{q}}$ and thereby does not include any damping. Since damping, in the form of Landau damping, occurs in even collisionless plasmas, this may be unrealistic. Damping also results from other dissipative processes, such as short range 'collisions' between particles. Damping can be artificially introduced into the model by generalizing the plasma modes to complex frequencies, by introducing the damping or 'collision frequency' $\nu_{\mathbf{q}}$ so that (85) becomes

$$\frac{1}{\varepsilon_{a}\left(\mathbf{q},\omega\right)}=1+\frac{\Omega_{a}^{\;2}}{\omega\left(\omega+\mathrm{i}\nu_{\mathbf{q}}^{a}\right)-\left(\Omega_{\mathbf{q}}^{a}\right)^{2}} \tag{89}$$

in which the collective mode now corresponds to the complex frequencies

$$\hat{\Omega}_{\mathbf{q}}^{a}=\pm\sqrt{\left(\Omega_{\mathbf{q}}^{a}\right)^{2}-\tfrac{1}{4}\left(\nu_{\mathbf{q}}^{a}\right)^{2}}-\tfrac{1}{2}\mathrm{i}\nu_{\mathbf{q}}^{a} \tag{90}$$

However this does not resolve the problem of Landau damping, which becomes exponentially small as $q\to 0$. In the classical limit of the RPA, for example [16], the damping is represented by the frequency

$$\nu_{\mathbf{q}}=\sqrt{\frac{\pi}{2}}\frac{\Omega}{\left(qD\right)^{3}}\exp\left[-\frac{1}{2\left(qD\right)^{2}}-\frac{3}{2}\right] \tag{91}$$

(see also section 4.2.6) so any physically reasonable model of Landau damping in the small $q$ regime would leave (88) essentially unmodified. This means that the collision frequency in (89) is generally interpreted as being that due to collisions, while the formula would not generally be considered to apply in the static limit.

Equation (89) is a generalization of the Drude formula, to which it reduces when $\Omega_{\mathbf{q}}^{a}=\Omega_{a}$.

### 3.2.1 Static Structure Factor

The static structure factor (44) is given, in the plasmon pole approximation, by

$$S_{\mathbf{q}}^{a} = q^2 D_a{}^2 \left(\frac{\Omega_a}{\Omega_{\mathbf{q}}^{a}}\right)^2 \frac{\Omega_{\mathbf{q}}^{a}}{2T} \coth \frac{\Omega_{\mathbf{q}}^{a}}{2T}$$

$$= q^2 D_a{}^2 \frac{\alpha_{\mathbf{q}}^{a}}{1+\alpha_{\mathbf{q}}^{a}} + \frac{q^2}{4m_a T}\left(\frac{2T}{\Omega_{\mathbf{q}}^{a}}\right)^2 \left(\frac{\Omega_{\mathbf{q}}^{a}}{2T}\coth\frac{\Omega_{\mathbf{q}}^{a}}{2T} - 1\right) \tag{92}$$

which vanishes in the limit $q \to 0$ for finite $\Omega_a$. However, taking the limit as the Coulomb interaction $e^2 \to 0$, yields, for the non-interacting limit

$$\lim_{e\to 0} S_{\mathbf{q}}^{a} = q^2 D_a{}^2 \left(\frac{\Omega_a}{\Omega_{\mathbf{q}}^{a}}\right)^2 \frac{\Omega_{\mathbf{q}}^{a}}{2T} \coth \frac{\Omega_{\mathbf{q}}^{a}}{2T}$$

$$= q^2 D_a{}^2 \alpha_{\mathbf{q}}^{a} + \frac{q^2}{12 m_a T} \tag{93}$$

$$= \tilde{R}_{\mathbf{q}}^{a} + \frac{q^2}{12 m_a T}$$

where $\tilde{R}_{\mathbf{q}}$ is defined, as before, by (42). Then, making use of (47),

$$\lim_{e\to 0} S_{\mathbf{q}}^{a} = R_{\mathbf{q}}^{a} + \mathcal{O}\left(q^4\right) \tag{94}$$

which therefore yields the non-interacting RPA static structure factor (63) to $\mathcal{O}\left(q^2\right)$. Equation (93) is exact in the context of the plasmon pole approximation and it follows from the sum rules that

$$\lim_{e\to 0} S_{\mathbf{q}}^{a} = \frac{2}{\pi} q^2 D_a{}^2 \int_0^\infty \left(1 + \frac{1}{3}\left(\frac{\omega}{2T}\right)^2\right) \operatorname{Im} \varepsilon_a\left(\mathbf{q},\omega\right) \frac{\mathrm{d}\omega}{\omega} \tag{95}$$

which should be compared with

$$\lim_{e\to 0} S_{\mathbf{q}}^{a} = \frac{2}{\pi} q^2 D_a{}^2 \int_0^\infty \frac{\omega}{2T} \coth\left(\frac{\omega}{2T}\right) \operatorname{Im} \varepsilon_a\left(\mathbf{q},\omega\right) \frac{\mathrm{d}\omega}{\omega} \tag{96}$$

which is the corresponding result in the RPA.

### 3.2.2 Multi-Component Plasmas in the Plasmon-pole Approximation

Because the simple plasmon-pole approximation (SPPA) does not encompass the RPA, except in the limit $q \to 0$, there is no justification for invoking (81). Instead, the plasmon pole approximation naturally generalizes to multi-component plasmas by means of a summation over many poles:

$$\frac{1}{\varepsilon(\mathbf{q},\omega)} = 1 + \sum_a \left( \frac{1}{\tilde{\varepsilon}_a(\mathbf{q},\omega)} - 1 \right) = 1 + \sum_a \frac{\Omega_a{}^2}{\omega\left(\omega + \mathrm{i}\tilde{\nu}_{\mathbf{q}}^a\right) - \left(\tilde{\Omega}_{\mathbf{q}}^a\right)^2} \tag{97}$$

where the resonance parameters, $\tilde{\Omega}_{\mathbf{q}}^a, \tilde{\nu}_{\mathbf{q}}^a$, represent the complete set of solutions of $\varepsilon\left(\mathbf{q}, \widehat{\Omega}_{\mathbf{q}}^a\right) = 0$, where $\widehat{\Omega}_{\mathbf{q}}^a$ is given by $\widehat{\Omega}_{\mathbf{q}}^a = \pm\sqrt{\left(\tilde{\Omega}_{\mathbf{q}}^a\right)^2 - \frac{1}{4}\left(\tilde{\nu}_{\mathbf{q}}^a\right)^2} - \frac{1}{2}\mathrm{i}\tilde{\nu}_{\mathbf{q}}^a$. Note that these resonance parameters will generally be shifted from the OCP values. The relation (82) is preserved by the above, but (83) and (84) are replaced by

$$\frac{\alpha_{\mathbf{q}}}{1+\alpha_{\mathbf{q}}} = \sum_a \frac{\tilde{\alpha}_{\mathbf{q}}^a}{1+\tilde{\alpha}_{\mathbf{q}}^a} = \sum_a \frac{\Omega_a{}^2}{\tilde{\Omega}_{\mathbf{q}}^{a2}} \tag{98}$$

where

$$\left(\tilde{\Omega}_{\mathbf{q}}^a\right)^2 = \Omega_a{}^2 \left(1 + \frac{1}{\tilde{\alpha}_{\mathbf{q}}^a}\right) \tag{99}$$

However, (98) implies that, unlike the RPA, in the presence of more than one component, $\tilde{\alpha}_{\mathbf{q}}^a$ is finite for $q \to 0$, which means that one has to be careful about the interpretations of $\tilde{\alpha}_{\mathbf{q}}^a$ and $\Omega_a$.

For single-component plasmas at least, the simple plasmon pole approximation has great appeal for its simplicity and tractability. Its main limitation is that it does not extend to the weak-coupling RPA regime and therefore cannot represent a universal model of a plasma. The reason for this is that it represents the system purely in terms of the collective modes, and makes no allowance for the possible presence of non-collective independent particle modes (for example). However it may provide a reasonable representation of some strongly-coupled plasmas in regimes where the dynamics are completely dominated by collective modes, eg so-called solid-state plasmas as well as solid state ionic lattices.

## 4 GENERALIZED PLASMON POLE APPROXIMATION

A generalization of the simple plasmon pole approximation that avoids these difficulties will now be presented.

Let $\varepsilon_a(\alpha,\omega)$ be a parametric representation of a one-body dielectric function such that

$$\varepsilon_a(\alpha_{\mathbf{q}}^a,\omega)=\varepsilon_a(\mathbf{q},\omega)$$
$$\varepsilon_a(\alpha,0)=1+\alpha \tag{100}$$

for $\alpha>0$, and define the generalization $\tilde{\varepsilon}_a(\alpha,\beta;\omega)$ by

$$\frac{1}{\tilde{\varepsilon}_a(\alpha,\beta;\omega)}=\frac{(1+\beta)^2}{\varepsilon_a(\alpha,\omega)+\beta} \tag{101}$$

where, for $\beta\geq 0$, $\tilde{\varepsilon}_a(\alpha,\beta;\omega)$ is free of zeros in the domain $\operatorname{Im}\omega>0$. Now suppose that, for some $F(\alpha,\beta)$ to be determined,

$$\operatorname{Im}\left(\frac{-1}{\tilde{\varepsilon}_a(\alpha,\beta;\omega)}\right)-F_a(\alpha,\beta)\operatorname{Im}\varepsilon_a(\alpha,\omega)=\frac{\pi}{2\omega}\Omega_a{}^2\left(1-F_a(\alpha,\beta)\right)h_{\alpha\beta}^a(\omega) \tag{102}$$

where, referring to the Cauchy identity and placing the poles in $\operatorname{Im}\omega<0$,

$$h_{\alpha\beta}^a(\omega)=\delta\left(\omega-\Omega_{\alpha\beta}^a\right)+\delta\left(\omega+\Omega_{\alpha\beta}^a\right)$$
$$=\frac{2}{\pi}\operatorname{Im}\frac{\omega}{\left(\Omega_{\alpha\beta}^a\right)^2-\left(\omega+\mathrm{i}0^+\right)^2} \tag{103}$$

Equation (102) is, by construction, consistent with the conductivity sum rule (31) and the generalized f-sum rule (33).

The generalization to damped modes can also be made at this stage by introducing the damping frequencies $\nu_{\alpha\beta}^a$ in the manner above, so that, for $\operatorname{Im}\omega=0$,

$$h_{\alpha\beta}^a(\omega)=\frac{2}{\pi}\operatorname{Im}\frac{\omega}{\left(\Omega_{\alpha\beta}^a\right)^2-\omega\left(\omega+\mathrm{i}\nu_{\alpha\beta}^a\right)} \tag{104}$$

The analytic continuation of (102) into the complex-$\omega$ domain is,

$$1+\beta-\frac{1}{\tilde{\varepsilon}_a(\alpha,\beta;z)}-F_a(\alpha,\beta)\left(\varepsilon_a(\alpha,z)-1\right)=\frac{1-F_a(\alpha,\beta)}{G_a(\alpha,\beta)-z} \tag{105}$$

where

$$z = \frac{\left(\omega + \mathrm{i}0^+\right)^2}{\Omega_a{}^2} \to \frac{\omega\left(\omega + \mathrm{i}\nu^a_{\alpha\beta}\right)}{\Omega_a{}^2} \tag{106}$$

and

$$G_a\left(\alpha,\beta\right) = \left(\frac{\Omega^a_{\alpha\beta}}{\Omega_a}\right)^2 \tag{107}$$

Equation (105) is the basic form of the Generalized Plasmon Pole approximation. It reduces to the simple plasmon pole approximation (89) when $F_a = 0,\ \ \beta = 0$.

Combining equations (101) and (105) yields

$$\left(\varepsilon_a - 1\right)\left(\frac{1+\beta}{\varepsilon_a + \beta} - F_a\right) = \frac{1 - F_a}{G_a - z} \tag{108}$$

another form of which is

$$\left(1 - F_a\right)\left(\varepsilon_a - 1 - \frac{1}{G_a - z}\right) = \frac{\left(\varepsilon_a - 1\right)^2}{\varepsilon_a + \beta} \tag{109}$$

These equations are postulated to hold at all frequencies and wavelengths. In particular, at zero frequency, $z = 0$, equation (109) implies

$$\left(1 - F_a\right)\left(\alpha - \frac{1}{G_a}\right) = \frac{\alpha^2}{1 + \alpha + \beta} \tag{110}$$

which, when solved for $F_a$ and $1 - F_a$, yields

$$\begin{aligned} 1 - F_a &= \left(\frac{\alpha G_a}{\alpha G_a - 1}\right)\left(\frac{\alpha}{1 + \alpha + \beta}\right) \\ F_a &= \left(1 - \frac{\alpha}{\left(\alpha G_a - 1\right)\left(1 + \beta\right)}\right)\left(\frac{1+\beta}{1 + \alpha + \beta}\right) \end{aligned} \tag{111}$$

from which it follows that $F_a$ is in the range $\left(0,1\right)$ if $G_a \geq \frac{1}{\alpha} + \frac{1}{1+\beta}$. Moreover, a further rearrangement of (110) gives

$$F_a + \frac{1 - F_a}{\alpha G_a} = \frac{1+\beta}{1 + \alpha + \beta} \tag{112}$$

which is in $(0,1)$ for $\alpha > 0, \quad \beta \geq 0$. With these relations in place, equation (105) is, in accordance with APPENDIX A, consistent with all the sum rules (31) - (34).

Now, from (108),

$$\frac{1}{z - G_a(\alpha,\beta)} = \frac{(\varepsilon_a - 1)}{1 - F_a(\alpha,\beta)}\left(F_a(\alpha,\beta) - \frac{1+\beta}{\varepsilon_a + \beta}\right) \tag{113}$$

We first consider this in the special case when $\beta = 0$, which applies to the physical case of a single plasma component. Let $G_a^0 = G_a(\alpha, 0)$ and $F_a^0 = F_a(\alpha, 0)$ in which case, using (111) - (112),

$$\begin{aligned} 1 - F_a^0 &= \left(\frac{\alpha G_a^0}{\alpha G_a^0 - 1}\right)\left(\frac{\alpha}{1+\alpha}\right) \\ F_a^0 &= \left(\frac{\alpha G_a^0 - 1 - \alpha}{\alpha G_a^0 - 1}\right)\left(\frac{1}{1+\alpha}\right) \end{aligned} \tag{114}$$

so that $F_a^0$ is found to be in $(0,1)$ if $G_a^0 > 1 + 1/\alpha$; and

$$F_a^0 + \frac{1 - F_a^0}{\alpha G_a} = \frac{1}{1+\alpha} \tag{115}$$

which is also in $(0,1)$ if $\alpha > 0$. In terms of these quantities, the general equation (113) takes the form

$$\frac{1}{z - G_a^0} = \frac{(\varepsilon_a - 1)}{1 - F_a^0}\left(F_a^0 - \frac{1}{\varepsilon_a}\right) \tag{116}$$

Now, for $\beta > 0$, the zeros of $\varepsilon_a + \beta$ in the frequency domain occur at $z = G_a$, whereupon (116) yields

$$\begin{aligned} \frac{1}{G_a - G_a^0} &= -\frac{(1+\beta)}{1 - F_a^0}\left(F_a^0 + \frac{1}{\beta}\right) \\ &= (1+\beta)\left(1 - \frac{1+\beta}{\beta\left(1 - F_a^0\right)}\right) \end{aligned} \tag{117}$$

Substituting for $\left(1 - F_a^0\right)$ from (114) and rearranging, yields the relation between $G_a = G_a(\alpha,\beta)$ and $G_a^0 = G_a(\alpha, 0)$ ,

$$G_a - \frac{1}{\alpha} - \frac{1}{1+\beta} = \left(G_a^0 - \frac{1}{\alpha} - 1\right)\left(1 + \frac{\alpha\beta}{\left(\alpha G_a^0 - 1 - \alpha\right) + (1+\alpha)\left(\alpha G_a^0 - 1\right)}\right) \tag{118}$$

which is $\geq 0$ if $G_a^0 \geq 1+\frac{1}{\alpha}$. Therefore, provided that the elementary function $G_a^0$ satisfies $G_a^0(\alpha) \geq 1+\frac{1}{\alpha}$, then

$$\begin{array}{ccc} G_a^0(\alpha) \geq 1+\dfrac{1}{\alpha} & \Rightarrow & G_a(\alpha,\beta) \geq \dfrac{1}{1+\beta}+\dfrac{1}{\alpha} \\ \Downarrow & & \Downarrow \\ 0 \leq F_a^0(\alpha) < 1 & & 0 \leq F_a(\alpha,\beta) < 1 \end{array} \tag{119}$$

Equation (118) expresses the modified resonant frequencies ($\beta \neq 0$) in terms of the elementary resonance frequencies associated with the individual plasma components ($\beta = 0$). Later, in section 4.1, it is shown how $z = G^0$ corresponds to the zeros of the physical dielectric functions of an individual component and therefore represents the resonances in the one-component system. The parameter $\beta$ thus expresses the shift in the resonance frequency due to the presence of other components, as is demonstrated in section 4.2.2.

## 4.1 Analytic Structure of the GPPA Dielectric Function for a One-Component Plasma

In the case of a single component ($\beta = 0$) equation (108) reduces to

$$(\varepsilon-1)(1-F\varepsilon) = (1-F)P\varepsilon \tag{120}$$

where

$$P = P(z) = \frac{1}{G-z} \tag{121}$$

Equation (120) is a quadratic equation for $\varepsilon$, whose two solutions shall be denoted $\varepsilon_+$ and $\varepsilon_-$. The following limiting cases arise: $F=1$ : $\varepsilon_+ = \varepsilon_- = 1$; and $F=0$: when the equation becomes linear with the single solution $\varepsilon = \frac{1}{1-P} = \frac{G-z}{G-z-1}$, which corresponds to the SPPA. Intermediate values of *F* therefore describe systems in regimes ranging from completely non-interacting to one dominated entirely by the resonant collective modes. Since $\lim_{\alpha\to 0} F(\alpha,0) = 1$ and $\lim_{\alpha\to\infty} F(\alpha,0) = 0$, it follows that the GPPA dielectric function, for a single component system, tends to unity (becomes weakly interacting) at small wavelengths ($\alpha \to 0$), and is dominated by collective motion at large wavelengths ($\alpha \to \infty$). This is physically reasonable.

In the generic case of $0 < F < 1$, (120) possesses two distinct solutions satisfying

$$\varepsilon_+ \varepsilon_- = \frac{1}{F} \tag{122}$$

and

$$\varepsilon_+ + \varepsilon_- = (P+1)F - (P-1) \tag{123}$$

so that the zeros of one solution correspond to the poles of the other. Now define

$$P_\pm = \frac{1 \pm \sqrt{F}}{1 \mp \sqrt{F}} \tag{124}$$

so that $P_+ > P_-$. Moreover

$$\begin{aligned} P_+ P_- &= 1 \\ P_+ + P_- &= 2\frac{1+F}{1-F} \end{aligned} \tag{125}$$

The solutions $\varepsilon_\pm$ can now be explicitly defined by

$$\begin{aligned} \varepsilon_\pm &= 1 + \frac{1-F}{2F}\left(1 - P \pm \sqrt{(P-P_+)(P-P_-)}\right) \\ &= 1 + \frac{1-F}{2F}\frac{1}{G-z}\left(G - z - 1 \pm \sqrt{1 - 2\frac{1+F}{1-F}(G-z) + (G-z)^2}\right) \end{aligned} \tag{126}$$

where the square root functions are defined so that $-\frac{\pi}{2} < \arg\sqrt{f} < \frac{\pi}{2}$. Evaluating

$$\varepsilon_+(z) = 1 + \frac{1-F}{2F}\frac{1}{G-z}\left(G - z - 1 + \sqrt{1 - 2\frac{1+F}{1-F}(G-z) + (G-z)^2}\right) \tag{127}$$

near $z = G$ yields

$$\varepsilon_+(z) = \frac{z-G}{1-F} + \mathcal{O}(z-G)^2 \tag{128}$$

so that $z = G$ corresponds to the zeros of $\varepsilon_+(z)$. Moreover, setting $z = 0$ in (127) and making use of (110) yields

$$\varepsilon_+(0) = 1 + \frac{1-F}{2FG}\left(G - 1 + \sqrt{\left(1 + \frac{2}{\alpha} - G\right)^2}\right) \tag{129}$$

and so, provided that

$$G \geq 1 + \frac{2}{\alpha} \tag{130}$$

and, with the aid of (111), one recovers the relation $\varepsilon_+(0) = 1 + \alpha$. The physical solution of (120) is therefore identified as $\varepsilon_+$, provided that $G$ satisfies the condition (130). The other solution represents an inverse dielectric function, and can be readily deduced from $\varepsilon_+$ using (122). However, if (130) is not satisfied, the physical solution becomes $\varepsilon_-$ (as in the SPPA) when $G$ is given by (88). This is a consequence of the definition of the selected Riemann branch of the square root function, which could, in principle, be modified so that $\varepsilon_+$ remains the physical branch. This appears to be a matter of mathematical convention, with no physical consequences. However, as we find later (in section 4.2.3) a violation of condition (130) by the ionic component can have adverse consequences for multicomponent plasmas. However, as we envisage only applications for which (130) holds, we do not need to worry about this.

The condition (130) is generally well satisfied by the RPA dispersion relation (79) except (for small $\mathbf{q}$) in cases of extreme electron degeneracy ($\gamma = 9/5$ in (79)) when the dielectric function is $\varepsilon_- = 1 + \dfrac{\alpha}{\gamma - 1}, \quad 2 \geq \gamma \geq 9/5$. The problem with multicomponent plasmas identified at equation (167) would require the ions be degenerate, so is never an issue.

The analytic properties of $\varepsilon_+(z) = \varepsilon_+\left(\omega(\omega + \mathrm{i}\nu)/\Omega^2\right)$ will now be considered in more detail. First of all, we note that, from the crossing relation (20), there is reflection symmetry about the imaginary-$\omega$ axis. It is therefore only necessary, in the first instance, to consider the function $\varepsilon_+$ in the domain $\mathrm{Re}\,\omega \geq 0$. In the $z$-plane, (126) contains branch points at

$$G - z = P_\pm = \frac{1 \pm \sqrt{F}}{1 \mp \sqrt{F}} \tag{131}$$

The location of these branch points depends on the magnitude of $(G - P_\pm)\Omega^2$ as follows:

1. $(G - P_+)\Omega^2 > \frac{1}{4}\nu^2$:

$$\omega = \sqrt{(G - P_\pm)\Omega^2 - \tfrac{1}{4}\nu^2} - \tfrac{1}{2}\mathrm{i}\nu \tag{132}$$

Since $G \geq P_\pm$, these branch points are located in $\mathrm{Im}\,\omega \leq 0$, as shown schematically in Figure 3.

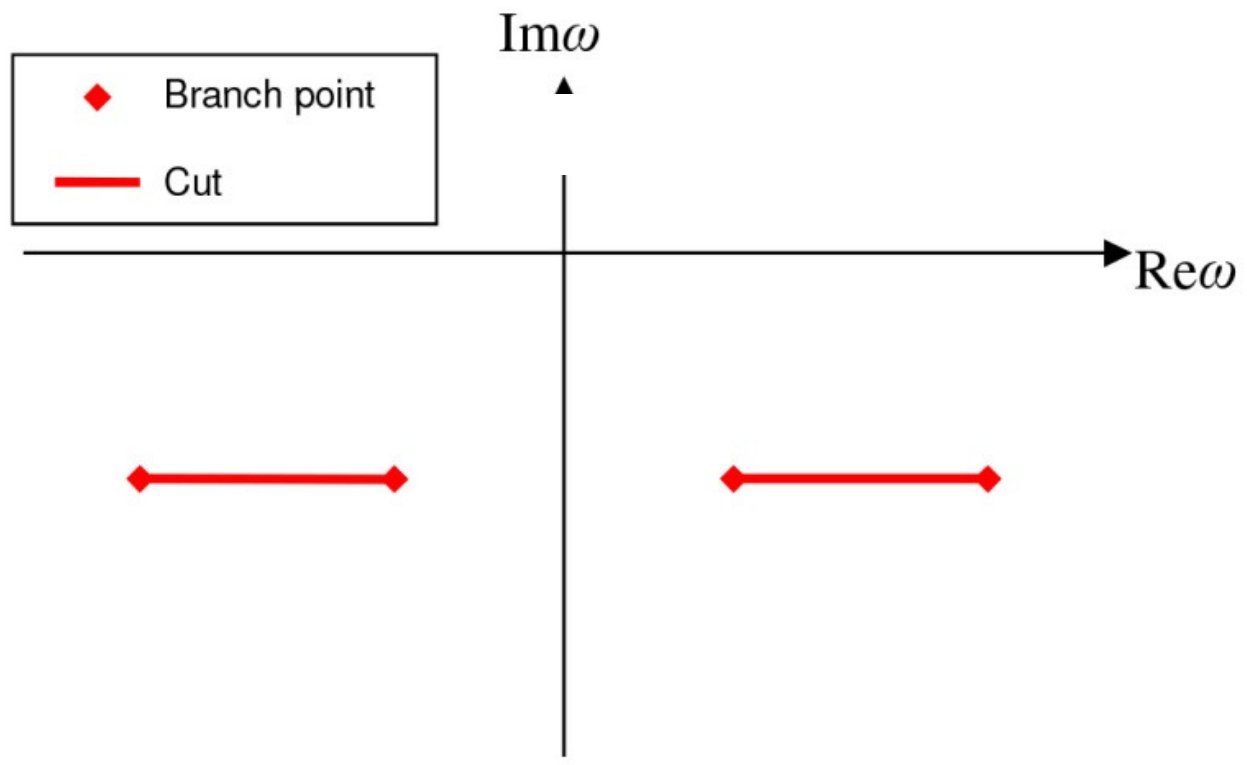

Figure 3: Complex $\omega$-plane showing showing the branch points of $\varepsilon(\omega)$, when $(G-P_+)\Omega^2 > \frac{1}{4}\nu^2$, as given by (132).

2. $0<(G-P_+)\Omega^2 \le \frac{1}{4}\nu^2, \quad (G-P_-)\Omega^2 > \frac{1}{4}\nu^2$:

$$\omega = -\mathrm{i}\left[\tfrac{1}{2}\nu \pm \sqrt{\tfrac{1}{4}\nu^2-(G-P_+)\Omega^2}\right], \quad -\tfrac{1}{2}\mathrm{i}\nu + \sqrt{(G-P_-)\Omega^2-\tfrac{1}{4}\nu^2} \tag{133}$$

Two of these branch points are on the negative imaginary axis, and the third and fourth are in $\mathrm{Im}\,\omega<0, \quad \mathrm{Re}\,\omega>0$ and $\mathrm{Im}\,\omega<0, \quad \mathrm{Re}\,\omega<0$, as shown in Figure 4.

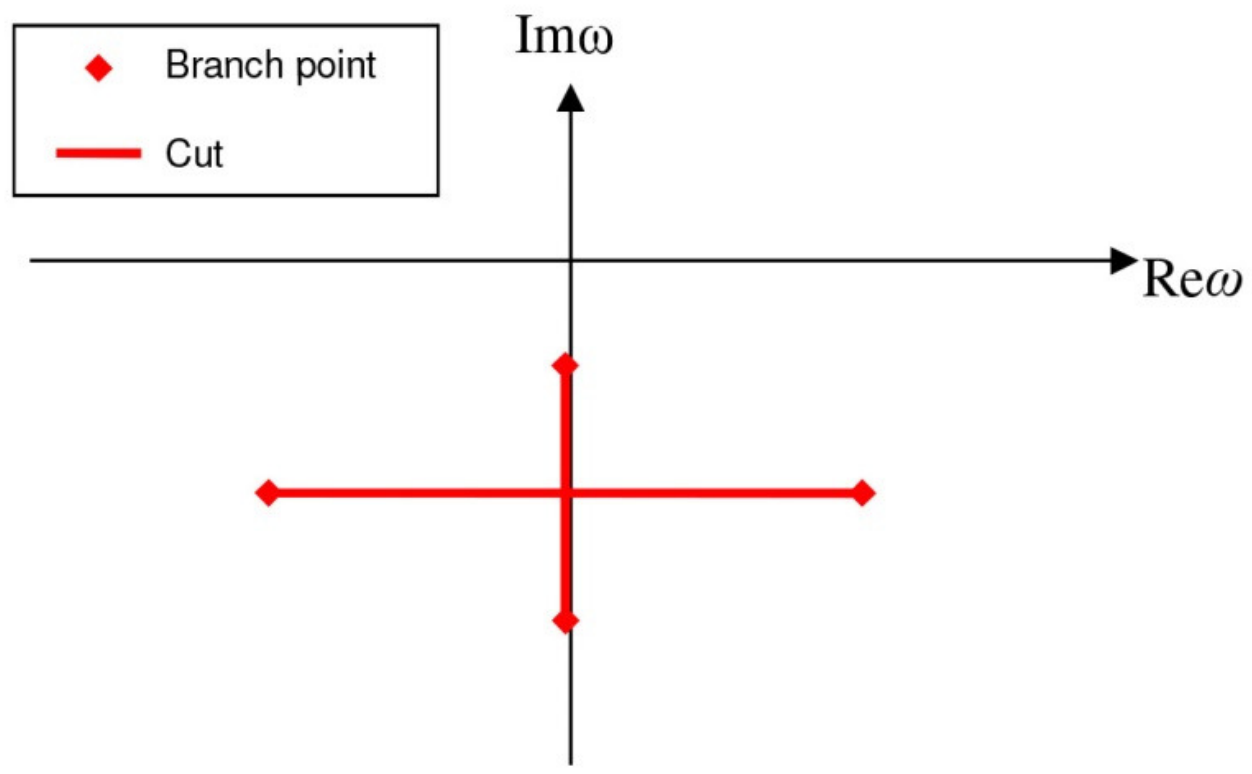

Figure 4: Complex $\omega$-plane showing showing the branch points of $\varepsilon(\omega)$ when $0<(G-P_+)\Omega^2 \le \frac{1}{4}\nu^2, \quad (G-P_-)\Omega^2 > \frac{1}{4}\nu^2$, as given by (133).

Note that the crossing relations imply $\mathrm{Re}\,\varepsilon\left(\omega-\frac{1}{2}\mathrm{i}\nu\right)=\mathrm{Re}\,\varepsilon\left(-\omega-\frac{1}{2}\mathrm{i}\nu\right)$ implying that the real part of $\varepsilon$ is continuous across the imaginary $\omega$-axis; and $\mathrm{Im}\,\varepsilon\left(\omega-\frac{1}{2}\mathrm{i}\nu\right)=-\mathrm{Im}\,\varepsilon\left(-\omega-\frac{1}{2}\mathrm{i}\nu\right)$ implying that the imaginary part of $\varepsilon$ changes sign across the imaginary axis. The cut on the negative imaginary axis therefore corresponds to $\varepsilon(\omega)$ having a non-zero imaginary part.

3. $0<\left(G-P_{\pm}\right)\Omega^2\le\frac{1}{4}\nu^2$ :

$$\omega=+\mathrm{i}\sqrt{\tfrac{1}{4}\nu^2-\left(G-P_{\pm}\right)\Omega^2}-\tfrac{1}{2}\mathrm{i}\nu,\quad -\mathrm{i}\sqrt{\tfrac{1}{4}\nu^2-\left(G-P_{\pm}\right)\Omega^2}-\tfrac{1}{2}\mathrm{i}\nu \tag{134}$$

All four branch points are now on the negative imaginary axis, with the two cuts corresponding to two regions on the negative imaginary axis where $\varepsilon$ is imaginary.

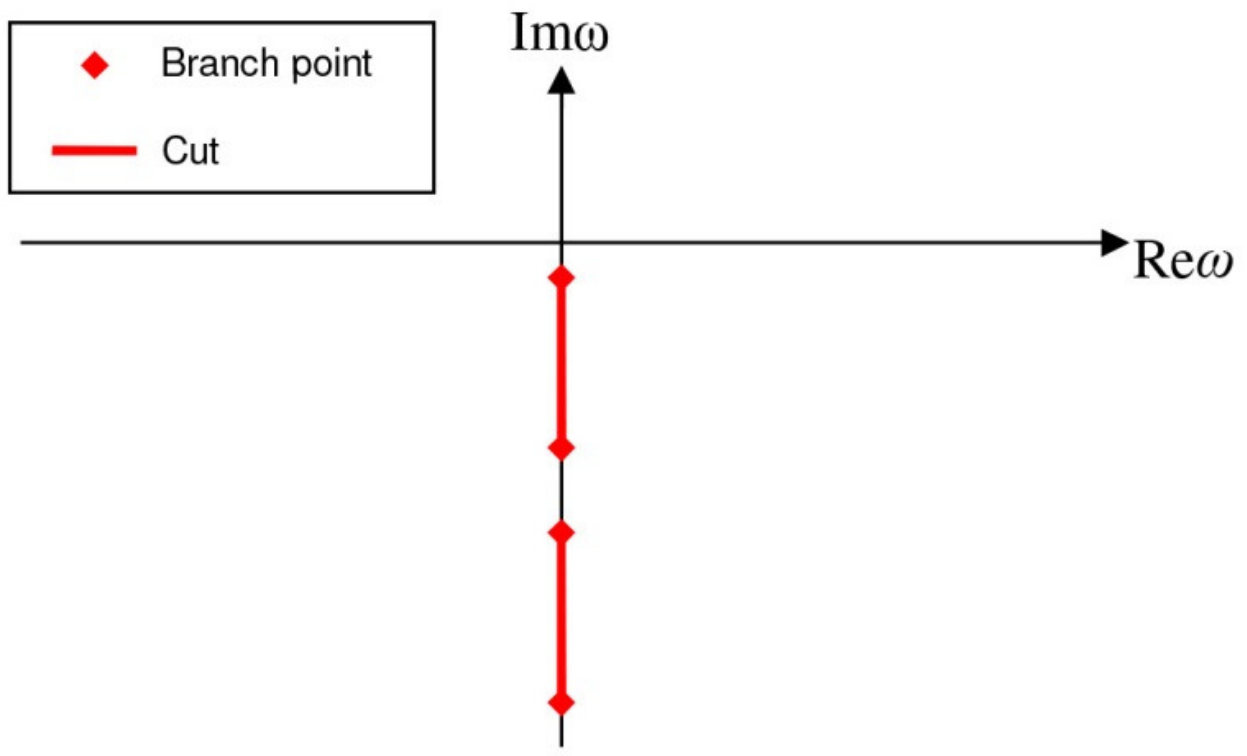

Figure 5: Complex $\omega$-plane showing showing the branch points of $\varepsilon_+(\omega)$, when $0<\left(G-P_{\pm}\right)\Omega^2\le\frac{1}{4}\nu^2$, as given by (134).

Thus to summarise, provided that $G>P_+$, the singularities of the physical dielectric function $\varepsilon_+$ correspond to branch points in $\mathrm{Im}\,\omega<0$. There are four branch points in all, either occurring in pairs on either side of the imaginary axis, one pair on the imaginary axis and another on either side of it, or two pairs both on the negative imaginary axis. In all cases, the Riemann cuts are confined to $\mathrm{Im}\le 0$. The function $\varepsilon_+(\omega)$ therefore has singularities only in the lower half plane, as required by causality. The function also has zeros wherever $z=G$, which are also, for $\nu>0$, confined to the lower half plane.

According to (119), the condition $G \geq P_+$ is implied by $G \geq 1+\frac{1}{\alpha}$, with $G = P_+$ when $G = 1+\frac{2}{\alpha}$, which follows from (129) in conjunction with (129). Therefore

$$G > 1+\frac{2}{\alpha} \tag{135}$$

is sufficient to ensure that the singularities are confined to $\operatorname{Im}\omega < 0$ in all cases. Generally, for not-too-degenerate plasmas, for which $\gamma \geq 2$, it is possible to get away with enforcing only the less strict (necessary) condition, $G \geq 1+1/\alpha$.

## 4.2 GPPA Treatment of Multicomponent Plasmas

The GPPA dielectric function for a multicomponent plasma is postulated as being given, by a generalization of (97), as follows:

$$\begin{aligned} \operatorname{Im}\frac{1}{\varepsilon(\mathbf{q},\omega)} &= \sum_a \lambda_a \operatorname{Im}\left(\frac{1}{\tilde{\varepsilon}_a(\alpha_a,\beta_a;\omega)}\right) \\ &= \sum_a \lambda_a \operatorname{Im}\left(\frac{(1+\beta_a)^2}{\varepsilon_a(\alpha_a,\omega)+\beta_a}\right) \end{aligned} \tag{136}$$

the analytic continuation of which is

$$\begin{aligned} 1-\frac{1}{\varepsilon(\mathbf{q},\omega)} &= \sum_a \lambda_a \left(1+\beta_a - \frac{(1+\beta_a)^2}{\varepsilon_a(\alpha_a,\omega)+\beta_a}\right) \\ &= \sum_a \frac{\lambda_a(\varepsilon_a - 1)(1+\beta_a)}{\varepsilon_a+\beta_a} \end{aligned} \tag{137}$$

where $\lambda_a$ and $\beta_a$ are parameters that do not depend on $\omega$ (but which, in general, do depend on $\mathbf{q}$).

Combining (136) and (137) with (105) yields

$$1-\frac{1}{\varepsilon(\mathbf{q},\omega)} = \sum_a \lambda_a \left( F_a(\alpha_a,\beta_a)(\varepsilon_a(\alpha_a,z)-1) + \frac{1-F_a(\alpha_a,\beta_a)}{G_a(\alpha_a,\beta_a)-z}\right) \tag{138}$$

$$\operatorname{Im}\left(\frac{-1}{\varepsilon(\mathbf{q},\omega)}\right) = \sum_a \lambda_a \left( F_a(\alpha_a,\beta_a)\operatorname{Im}\varepsilon_a(\alpha_a,z) + (1-F_a(\alpha_a,\beta_a))\operatorname{Im}\left(\frac{1}{G_a - z}\right)\right) \tag{139}$$

Equations (138) and (139) are the GPPA equations for a multicomponent plasma. The right hand side of (139) is a sum over the components in which each term comprises a pole contribution $\propto \mathrm{Im}(G_a - z)^{-1}$ and a 'background' term $\propto \mathrm{Im}\,\varepsilon_a(\alpha_a, z)$. The coefficients of each pair of terms lie between 0 and 1 and add up to unity, representing a sharing of strength between the two parts. This is best illustrated by evaluating the f-sum rule (38) with (139), which yields

$$\int_0^\infty \omega \mathrm{Im}\left(\frac{-1}{\varepsilon(\mathbf{q},\omega)}\right)\mathrm{d}\omega = \frac{\pi}{2}\sum_a \lambda_a {\Omega_a}^2 \tag{140}$$

so that, in this approximation, the limiting plasma frequency is given by $\Omega^2 = \sum_a \lambda_a {\Omega_a}^2$. In general, the $\Omega_a$ should be considered as representing the normal modes of the plasma, rather than the constituent modes.

In situations when the resonance modes are weakly damped

$$\mathrm{Im}\left(\frac{1}{G_a - z}\right) = \frac{\pi {\Omega_a}^2}{2\omega}\left(\delta\left(\omega - \Omega_{\mathbf{q}}^a\right) + \delta\left(\omega + \Omega_{\mathbf{q}}^a\right)\right) \tag{141}$$

where

$$\left(\Omega_{\mathbf{q}}^a\right)^2 = G_a(\alpha_a, \beta_a){\Omega_a}^2 \tag{142}$$

are the resonant frequencies of the plasma mixture. Equation (139) represents a modal decomposition of the dielectric function into particle-like modes represented by $\varepsilon_a(\alpha_a, z)$ and collective resonances represented by the poles, $(G_a - z)^{-1}$.

#### 4.2.1 Random Phase Approximation

Unlike the SPPA, the GPPA is sufficiently general to encompass the RPA, provided that

$$\sum_a \lambda_a {\Omega_a}^2 = \sum_a {\Omega_a}^2 \tag{143}$$

which results from aligning the RPA f-sum rule with (140); together with

$$\left(\sum_a \alpha_a\right)\left(1 - \sum_a \frac{\lambda_a \alpha_a (1+\beta_a)}{1+\alpha_a+\beta_a}\right) = \sum_a \frac{\lambda_a \alpha_a (1+\beta_a)}{1+\alpha_a+\beta_a} \tag{144}$$

so that, from (137),

$$\varepsilon(\mathbf{q},0) - 1 \equiv \alpha = \sum_a \alpha_a \tag{145}$$

### 4.2.2 Two Component Plasma in the RPA

To see how this works, consider a two component plasma (TCP) consisting of a heavy particle species (ions, $a=\mathrm{i}$) and a light species (electrons, $a=\mathrm{e}$).

Then, using the RPA formula (81) and assuming $m_\mathrm{e} << m_\mathrm{i}$, we have

$$\mathrm{Im}\left(\frac{-1}{\varepsilon(\mathbf{q},\omega)}\right)=\frac{\mathrm{Im}\,\varepsilon_\mathrm{i}(\mathbf{q},\omega)}{|\varepsilon(\mathbf{q},\omega)|^2}+\frac{\mathrm{Im}\,\varepsilon_\mathrm{e}(\mathbf{q},\omega)}{|\varepsilon(\mathbf{q},\omega)|^2}$$

$$\simeq \mathrm{Im}\left(\frac{-1}{\varepsilon_\mathrm{i}(\mathbf{q},\omega)+\alpha_\mathrm{e}}\right)+\lambda_\mathrm{e}\,\mathrm{Im}\left(\frac{-1}{\varepsilon_\mathrm{e}(\mathbf{q},\omega)}\right) \tag{146}$$

$$=\frac{1}{(1+\alpha_\mathrm{e})^2}\mathrm{Im}\left(-\frac{(1+\alpha_\mathrm{e})^2}{\varepsilon_\mathrm{i}(\mathbf{q},\omega)+\alpha_\mathrm{e}}\right)+\lambda_\mathrm{e}\,\mathrm{Im}\left(\frac{-1}{\varepsilon_\mathrm{e}(\mathbf{q},\omega)}\right)$$

which is of the form (136), and complies with (143), provided that the coefficients $\beta_{\mathrm{e,i}}$ and $\lambda_{\mathrm{e,i}}$ are chosen as follows:

$$\begin{aligned}
\beta_\mathrm{i}&=\alpha_\mathrm{e}\\
\beta_\mathrm{e}&=0\\
\lambda_\mathrm{i}&=\frac{1}{(1+\alpha_\mathrm{e})^2}\\
\lambda_\mathrm{e}&=1+\left(1-\frac{1}{(1+\alpha_\mathrm{e})^2}\right)\left(\frac{\Omega_\mathrm{i}}{\Omega_\mathrm{e}}\right)^2
\end{aligned} \tag{147}$$

Making the approximation $\lambda_\mathrm{e}\approx 1$ ($m_\mathrm{e} << m_\mathrm{i}$) yields that

$$\begin{aligned}
\sum_a\frac{\lambda_a\alpha_a(1+\beta_a)}{1+\alpha_a+\beta_a}&=\frac{\alpha_\mathrm{e}}{1+\alpha_\mathrm{e}}+\frac{\alpha_\mathrm{i}(1+\alpha_\mathrm{e})}{(1+\alpha_\mathrm{e})^2(1+\alpha_\mathrm{i}+\alpha_\mathrm{e})}\\
&=\frac{\alpha_\mathrm{i}+\alpha_\mathrm{e}}{1+\alpha_\mathrm{i}+\alpha_\mathrm{e}}
\end{aligned} \tag{148}$$

which thereby satisfies the condition (144), and hence

$$\alpha=\alpha_\mathrm{i}+\alpha_\mathrm{e} \tag{149}$$

Thus the GPPA embodies facets of the RPA description of the two-component plasma consisting of electrons and ions.

#### 4.2.3 The GPPA Dynamic Structure Factor

The charge-density dynamic structure factor, which is the Fourier transform of the space and time charge density correlation function, is determined, in the modal decomposition, from (139), by the fluctuation-dissipation theorem (26), as follows

$$S(\mathbf{q},\omega)=\sum_a \lambda_a w_a\left(F_a(\alpha_a,\beta_a)S_a^0(\mathbf{q},\omega)+\left(1-F_a(\alpha_a,\beta_a)\right)\frac{q^2}{2m_a}g_{\mathbf{q}}^a(\omega)\right) \tag{150}$$

where

$$w_a=\frac{m_a{\Omega_a}^2}{m_\mathrm{e}{\Omega_\mathrm{e}}^2} \tag{151}$$

which is the charge density weighting;

$$\begin{aligned} S_a^0(\mathbf{q},\omega)&=\left|\varepsilon_a(\mathbf{q},\omega)\right|^2 S_a(\mathbf{q},\omega)\\ &=\frac{q^2}{\pi m_a{\Omega_a}^2}\frac{1}{1-\mathrm{e}^{-\omega/T}}\operatorname{Im}\varepsilon_a(\mathbf{q},\omega) \end{aligned} \tag{152}$$

in which $S_a(\mathbf{q},\omega)$ is the one-component structure factor for charge species a :

$$S_a(\mathbf{q},\omega)=\frac{q^2}{\pi m_a{\Omega_a}^2}\frac{1}{1-\mathrm{e}^{-\omega/T}}\operatorname{Im}\left(\frac{-1}{\varepsilon_a(\mathbf{q},\omega)}\right) \tag{153}$$

and

$$g_{\mathbf{q}}^a(\omega)=\frac{2}{\pi{\Omega_a}^2}\frac{1}{1-\mathrm{e}^{-\omega/T}}\operatorname{Im}\left(\frac{1}{G_a-z}\right) \tag{154}$$

If $\varepsilon_a(\mathbf{q},\omega)$ is given in the RPA, then $S_a^0(\mathbf{q},\omega)$ is the non-interacting structure factor for species *a.*

The quantities defined by (152) - (154) satisfy the following integral relations,

$$\begin{aligned} \int_{-\infty}^{\infty} S_a^0(\mathbf{q},\omega)\,\mathrm{d}\omega&=\frac{q^2}{\pi m_a{\Omega_a}^2}\int_{-\infty}^{\infty}\frac{1}{1-\mathrm{e}^{-\omega/T}}\operatorname{Im}\varepsilon_a(\mathbf{q},\omega)\,\mathrm{d}\omega\\ &=\frac{2}{\pi}q^2{D_a}^2\int_0^{\infty}\frac{\omega}{2T}\coth\left(\frac{\omega}{2T}\right)\operatorname{Im}\varepsilon_a(\mathbf{q},\omega)\frac{\mathrm{d}\omega}{\omega}\\ &=R_{\mathbf{q}}^a \end{aligned} \tag{155}$$

where $R^a_{\mathbf{q}}$ is defined, for species a, by (43);

$$\int_{-\infty}^{\infty} S_a^0(\mathbf{q},\omega)\,\omega\,\mathrm{d}\omega = \frac{q^2}{\pi m_a \Omega_a{}^2}\int_{-\infty}^{\infty} \operatorname{Im}\varepsilon_a(\mathbf{q},\omega)\,\omega\,\mathrm{d}\omega$$
$$= \frac{q^2}{2m_a} \tag{156}$$

which follows directly from (20) and (31);

$$\int_{-\infty}^{\infty} S_a(\mathbf{q},\omega)\,\mathrm{d}\omega = \frac{2}{\pi} q^2 D_a{}^2 \int_0^{\infty} \frac{\omega}{2T}\coth\left(\frac{\omega}{2T}\right)\operatorname{Im}\left(\frac{-1}{\varepsilon_a(\mathbf{q},\omega)}\right)\frac{\mathrm{d}\omega}{\omega}$$
$$= S^a_{\mathbf{q}} \tag{157}$$

in accordance with (44);

$$\int_{-\infty}^{\infty} S_a(\mathbf{q},\omega)\,\omega\,\mathrm{d}\omega = \frac{q^2}{\pi m_a \Omega_a{}^2}\int_{-\infty}^{\infty} \operatorname{Im}\left(\frac{-1}{\varepsilon_a(\mathbf{q},\omega)}\right)\omega\,\mathrm{d}\omega$$
$$= \frac{q^2}{2m_a} \tag{158}$$

which follows directly from (20) and (38); and

$$\int_{-\infty}^{\infty} g^a_{\mathbf{q}}(\omega)\,\mathrm{d}\omega = \frac{2}{\pi}\int_{-\infty}^{\infty} \frac{1}{1-\mathrm{e}^{-\omega/T}}\operatorname{Im}\left(\frac{1}{\left(\Omega^a_{\mathbf{q}}\right)^2 - \omega\left(\omega + \mathrm{i}\nu^a_{\mathbf{q}}\right)}\right)\mathrm{d}\omega$$
$$= \frac{1}{\hat{\Omega}^a_{\mathbf{q}}}\coth\frac{\hat{\Omega}^a_{\mathbf{q}}}{2T}\left(1 + \sin^2\frac{\nu^a_{\mathbf{q}}}{4T}\Big/\sinh^2\frac{\hat{\Omega}^a_{\mathbf{q}}}{2T}\right)^{-1} \tag{159}$$
$$\simeq \frac{1}{\Omega^a_{\mathbf{q}}}\coth\frac{\Omega^a_{\mathbf{q}}}{2T}$$

in which

$$\left(\hat{\Omega}^a_{\mathbf{q}}\right)^2 = \left(\Omega^a_{\mathbf{q}}\right)^2 - \tfrac{1}{4}\left(\nu^a_{\mathbf{q}}\right)^2 \tag{160}$$

(cf. equation (90)). The approximation (159) is valid for all positive real values of $\hat{\Omega}^a_{\mathbf{q}}/T$ provided only that $\nu^a_{\mathbf{q}}/4T \lesssim 1$. The proof of the result (159)-(160) is given in Appendix B.3. Note that, in temperature regimes where there is appreciable excitation of collective modes, ie $T \gtrsim \Omega^a_{\mathbf{q}}$, this condition is implied by $\nu_{\mathbf{q}}^{a} \lesssim \Omega_{\mathbf{q}}^{a}$. The condition therefore amounts to a weak restriction. (See also section 4.2.6.)

Let

$$S_{a}^{0}(\mathbf{q},\omega) = R_{\mathbf{q}}^{a}\delta_{\mathbf{q}}^{a}(\omega) \tag{161}$$

where $R_{\mathbf{q}}^{a}$ is defined above, for any species, $a$, by (63), while, in accordance with (155) and (156), $\delta_{\mathbf{q}}^{a}(\omega)$ satisfies,

$$\begin{gathered}\int_{-\infty}^{\infty}\delta_{\mathbf{q}}^{a}(\omega)\,\mathrm{d}\omega = 1\\ \int_{-\infty}^{\infty}\delta_{\mathbf{q}}^{a}(\omega)\,\omega\,\mathrm{d}\omega = \frac{q^{2}}{2m_{a}R_{\mathbf{q}}^{a}}\end{gathered} \tag{162}$$

and let the charge-density dynamic structure factor be expressed in the form

$$S(\mathbf{q},\omega) = \sum_{a}\lambda_{a}\left(A_{\mathbf{q}}^{a}\delta_{\mathbf{q}}^{a}(\omega) + B_{\mathbf{q}}^{a}\frac{q^{2}}{2m_{a}}g_{\mathbf{q}}^{a}(\omega)\right) \tag{163}$$

Comparison with the above yields the coefficients $A_{\mathbf{q}}^{a}$ and $B_{\mathbf{q}}^{a}$ as follows:

$$\begin{gathered}A_{\mathbf{q}}^{a} = w_{a}R_{\mathbf{q}}^{a}F_{a}\left(\alpha_{\mathbf{q}}^{a},\beta_{a}\right)\\ B_{\mathbf{q}}^{a} = w_{a}\left(1 - F_{a}\left(\alpha_{\mathbf{q}}^{a},\beta_{a}\right)\right)\end{gathered} \tag{164}$$

which decomposes the dynamic structure factor into a central (Rayleigh) peak characterised by the $\delta_{\mathbf{q}}^{a}(\omega)$ terms and shifted resonance (Brillouin) lines characterised by the $g_{\mathbf{q}}^{a}(\omega)$ terms , where the coefficients $A_{\mathbf{q}}^{a}$ and $B_{\mathbf{q}}^{a}$ satisfy $A_{\mathbf{q}}^{a}/R_{\mathbf{q}}^{a} + B_{\mathbf{q}}^{a} = w_{a}$ . Equation (163) represents the modal decomposition of the dynamic structure factor. In particular, in the case of a two-component plasma,

$$\begin{gathered}A_{\mathbf{q}}^{\mathrm{e}} = R_{\mathbf{q}}^{\mathrm{e}}\frac{\alpha_{\mathrm{e}}G_{\mathrm{e}} - 1 - \alpha_{\mathrm{e}}}{(\alpha_{\mathrm{e}}G_{\mathrm{e}} - 1)(1 + \alpha_{\mathrm{e}})}\\ B_{\mathbf{q}}^{\mathrm{e}} = \left(\frac{\alpha_{\mathrm{e}}G_{\mathrm{e}}}{\alpha_{\mathrm{e}}G_{\mathrm{e}} - 1}\right)\left(\frac{\alpha_{\mathrm{e}}}{1 + \alpha_{\mathrm{e}}}\right)\end{gathered} \tag{165}$$

$$A_{\mathbf{q}}^{\mathrm{i}} = w_{\mathrm{i}}\left(1-\left(\frac{\alpha_{\mathrm{i}}G_{\mathrm{i}}}{\alpha_{\mathrm{i}}G_{\mathrm{i}}-1}\right)\left(\frac{\alpha_{\mathrm{i}}}{1+\alpha}\right)\right)$$

$$B_{\mathbf{q}}^{\mathrm{i}} = w_{\mathrm{i}}\left(\frac{\alpha_{\mathrm{i}}G_{\mathrm{i}}}{\alpha_{\mathrm{i}}G_{\mathrm{i}}-1}\right)\left(\frac{\alpha_{\mathrm{i}}}{1+\alpha}\right) \tag{166}$$

In the above, the functions $G_{\mathrm{i}} \equiv G_{\mathrm{i}}(\alpha_{\mathrm{i}}, \beta_{\mathrm{i}})$ and $G_{\mathrm{e}} \equiv G_{\mathrm{e}}(\alpha_{\mathrm{e}}, \beta_{\mathrm{e}})$ are given, by equation (118), in terms of the elementary functions $G_{\mathrm{i,e}}^{0}(\alpha)$, and the parameters $\alpha_{\mathrm{i,e}} = \varepsilon_{\mathrm{i,e}}(\mathbf{q},0)-1$ and $\beta_{\mathrm{i,e}}$ given by (147). Using the parameterisation (79) for the $G_{\mathrm{i,e}}^{0}(\alpha)$, this yields

$$G_{\mathrm{e}} = 1+\frac{\gamma_{\mathrm{e}}}{\alpha_{\mathrm{e}}}$$

$$G_{\mathrm{i}} = \frac{1}{1+\alpha_{\mathrm{e}}}+\frac{\gamma_{\mathrm{i}}}{\alpha_{\mathrm{i}}}+\frac{(\gamma_{\mathrm{i}}-1)\alpha_{\mathrm{e}}}{(\alpha_{\mathrm{i}}+2)(\alpha_{\mathrm{i}}+\gamma_{\mathrm{i}}-2)} \tag{167}$$

in which $\gamma_{\mathrm{e,i}} > 2$ as per (135). For regimes approximately described by the RPA, the assignments $\gamma_{\mathrm{i}} = \gamma_{\mathrm{e}} = 3$ are proposed, in accordance with (80), as a reasonable choice, at least within a limited range of wavenumbers. Equation (167) shows that $G_{\mathrm{i}}$ is singular for some finite positive value of $\alpha_{\mathrm{i}}$ (corresponding to finite *q*) if $1 < \gamma_{\mathrm{i}} < 2$, while $\gamma_{\mathrm{i}} = 1$ (SPPA) is an isolated special case.

#### 4.2.4 The GPPA Static Structure Factor

The static structure factor can now be calculated from (150), using (155) and (159), to give

$$\begin{aligned} S(\mathbf{q}) &= \int_{-\infty}^{+\infty} S(\mathbf{q},\omega)\,\mathrm{d}\omega \\ &= \sum_a \lambda_a w_a\left(F_a(\alpha_a,\beta_a)R_{\mathbf{q}}^a+\left(1-F_a(\alpha_a,\beta_a)\right)\frac{q^2}{2m_a\Omega_{\mathbf{q}}^a}\coth\frac{\Omega_{\mathbf{q}}^a}{2T}\right) \\ &= \sum_a \lambda_a w_a\left(F_a(\alpha_a,\beta_a)R_{\mathbf{q}}^a+\left(1-F_a(\alpha_a,\beta_a)\right)\frac{q^2 {D_a}^2}{G_a}\frac{\Omega_{\mathbf{q}}^a}{2T}\coth\frac{\Omega_{\mathbf{q}}^a}{2T}\right) \\ &= \sum_a \lambda_a w_a\left(F_a(\alpha_a,\beta_a)R_{\mathbf{q}}^a+\left(1-F_a(\alpha_a,\beta_a)\right)\frac{\tilde{R}_{\mathbf{q}}^a}{\alpha_a G_a}\frac{\Omega_{\mathbf{q}}^a}{2T}\coth\frac{\Omega_{\mathbf{q}}^a}{2T}\right) \end{aligned} \tag{168}$$

where use has been made of (107), (41), and (46). A rearrangement of terms in (168), using (111)-(112), yields

$$S(\mathbf{q}) = \sum_a \lambda_a w_a \left( \begin{array}{l} F_a(\alpha_a,\beta_a)\left(R_{\mathbf{q}}^a - \tilde{R}_{\mathbf{q}}^a\right) \\ +\tilde{R}_{\mathbf{q}}^a \left( F_a(\alpha_a,\beta_a) + \frac{1-F_a(\alpha_a,\beta_a)}{\alpha_a G_a} + \frac{1-F_a(\alpha_a,\beta_a)}{\alpha_a G_a}\left(\frac{\Omega_{\mathbf{q}}^a}{2T}\coth\frac{\Omega_{\mathbf{q}}^a}{2T} - 1\right)\right) \end{array} \right)$$

$$= \sum_a \lambda_a w_a \left( F_a(\alpha_a,\beta_a)\left(R_{\mathbf{q}}^a - \tilde{R}_{\mathbf{q}}^a\right) + \tilde{R}_{\mathbf{q}}^a \left( \frac{1+\beta_a}{1+\alpha_a+\beta_a} + \frac{\alpha_a}{(1+\alpha_a+\beta_a)(\alpha_a G_a - 1)}\left(\frac{\Omega_{\mathbf{q}}^a}{2T}\coth\frac{\Omega_{\mathbf{q}}^a}{2T} - 1\right)\right)\right)$$

$$= S^{(0)}(\mathbf{q}) + S^{(1)}(\mathbf{q}) + S^{(2)}(\mathbf{q}) + S^{(3)}(\mathbf{q}) \tag{169}$$

where

$$S^{(0)}(\mathbf{q}) = \sum_a \lambda_a w_a \tilde{R}_{\mathbf{q}}^a \left( F_a(\alpha_a,\beta_a) + \frac{1-F_a(\alpha_a,\beta_a)}{\alpha_a G_a} \right)$$

$$= \sum_a \lambda_a w_a \tilde{R}_{\mathbf{q}}^a \left( \frac{1+\beta_a}{1+\alpha_a+\beta_a} \right) \tag{170}$$

which is the leading classical term describing the non-resonant response;

$$S^{(1)}(\mathbf{q}) = \sum_a \lambda_a w_a \tilde{R}_{\mathbf{q}}^a \frac{1-F_a(\alpha_a,\beta_a)}{\alpha_a G_a}\left(\frac{\Omega_{\mathbf{q}}^a}{2T}\coth\frac{\Omega_{\mathbf{q}}^a}{2T} - 1\right)$$

$$= \sum_a \lambda_a w_a \tilde{R}_{\mathbf{q}}^a \frac{\alpha_a}{(1+\alpha_a+\beta_a)(\alpha_a G_a - 1)}\left(\frac{\Omega_{\mathbf{q}}^a}{2T}\coth\frac{\Omega_{\mathbf{q}}^a}{2T} - 1\right) \tag{171}$$

which is the leading (semi-classical) term describing the resonant response;

$$S^{(2)}(\mathbf{q}) = \sum_a \lambda_a w_a \left(R_{\mathbf{q}}^a - \tilde{R}_{\mathbf{q}}^a\right) \frac{1+\beta_a}{1+\alpha_a+\beta_a} \tag{172}$$

and

$$S^{(3)}(\mathbf{q}) = \sum_a \lambda_a w_a \left(R_{\mathbf{q}}^a - \tilde{R}_{\mathbf{q}}^a\right)\left( F_a(\alpha_a,\beta_a) - \frac{1+\beta_a}{1+\alpha_a+\beta_a} \right) \tag{173}$$

which represents additional quantum corrections. The terms $S^{(1)}$ , $S^{(2)}$ and $S^{(3)}$ vanish in the classical and high-temperature limits.

We consider these equations in a form that generalizes to systems in which the components are not in mutual thermal equilibrium, but are characterised by different temperatures, $T_a$, $T_b$…, so that the particle-particle pair correlation function $g(a|b)$, corresponding to the pair distribution function of particles of species *a* with respect to a particle of species *b*, $\rho(a|b) = n_a(1+g(a|b))$, satisfies the canonical (classical) relation

$$\frac{g(a|b)}{g(b|a)} = \frac{T_b}{T_a} \tag{174}$$

Bearing in mind that the structure factor is defined with respect to the charge density, and contains a normalisation factor of $1/n_e$, the pair correlation function in coordinate space is

$$n_a Z_a g_{\mathbf{r}}(a|b) = \frac{Z_b}{w_b}\frac{1}{(2\pi)^3}\int\left(S_{ab}(\mathbf{q}) - \delta_{ab}S_{ab}(\infty)\right)\mathrm{e}^{-\mathrm{i}\mathbf{q}\cdot\mathbf{r}}\mathrm{d}^3\mathbf{q} \tag{175}$$

which generalizes the OCP formula, and where the definition (151) of the coefficient $w_a$ is now generalized to

$$w_a = \frac{m_a \Omega_a{}^2 T_e}{m_e \Omega_e{}^2 T_a} = \frac{D_e{}^2}{D_a{}^2} \tag{176}$$

Note that it is important to pay attention to the relative signs of the charges $Z_a$ and $Z_b$ in equation (175). Then, using (41), (46) and (151), and replacing $\alpha_a$ by the equivalent quantities $\alpha_{\mathbf{q}}^a$ to reflect the hitherto suppressed dependence on $\mathbf{q}$,

$$\frac{\alpha_{\mathbf{q}}^a}{\alpha_{\mathbf{q}}^e} = \frac{\tilde{R}_{\mathbf{q}}^a}{\tilde{R}_{\mathbf{q}}^e} w_a \tag{177}$$

Substituting accordingly for $\tilde{R}_{\mathbf{q}}^a w_a$ in $S^{(0)}(\mathbf{q})$, and making use of (137), yields

$$\begin{aligned} S^{(0)}(\mathbf{q}) &= \frac{\tilde{R}_{\mathbf{q}}^e}{\alpha_{\mathbf{q}}^e}\sum_a \lambda_a \alpha_{\mathbf{q}}^a \left(\frac{1+\beta_a}{1+\alpha_{\mathbf{q}}^a+\beta_a}\right) \\ &= \frac{\tilde{R}_{\mathbf{q}}^e}{\alpha_{\mathbf{q}}^e}\left(1 - \frac{1}{\varepsilon(\mathbf{q},0)}\right) \\ &= q^2 D_e{}^2\left(1 - \frac{1}{\varepsilon(\mathbf{q},0)}\right) \end{aligned} \tag{178}$$

where use has been made of (137). The final form of the result (178) represents the simplest expression of the structure factor component $S^{(0)}(\mathbf{q})$ which is thus shown to be essentially classical in nature (although quantum mechanical terms are implicitly retained in $\varepsilon(\mathbf{q},0)$). We now consider the decomposition of this into its component two-particle correlations.

Applying the RPA decomposition of the dielectric function at zero frequency,

$$1-\frac{1}{\epsilon(\mathbf{q},0)}=\frac{\alpha_{\mathbf{q}}}{1+\alpha_{\mathbf{q}}}=\sum_a\frac{\alpha_{\mathbf{q}}^a}{1+\alpha_{\mathbf{q}}} \tag{179}$$

whereupon, from (178), making use of (177),

$$S^{(0)}(\mathbf{q})=\frac{\tilde{R}_{\mathbf{q}}^{\mathrm{e}}}{\alpha_{\mathbf{q}}^{\mathrm{e}}}\sum_a\frac{\alpha_{\mathbf{q}}^a}{1+\alpha_{\mathbf{q}}}=\sum_a\frac{w_a\tilde{R}_{\mathbf{q}}^a}{\varepsilon(\mathbf{q},0)} \tag{180}$$

from which one can determine the canonical classical decomposition of the total classical static structure factor into component particle terms representing the particle-particle correlations, as follows: First of all, substituting for $1/\varepsilon(\mathbf{q},0)$ according to the RPA identity relation,

$$\frac{1}{\varepsilon(\mathbf{q},0)}\equiv\sum_b\left(\delta_{ab}-\frac{\alpha_b}{\varepsilon(\mathbf{q},0)}\right) \tag{181}$$

yields

$$\begin{aligned}S^{(0)}(\mathbf{q})&=\sum_{a,b}w_a\tilde{R}_{\mathbf{q}}^a\left(\delta_{ab}-\frac{\alpha_b}{\varepsilon(\mathbf{q},0)}\right)\\&=\sum_{a,b}w_a\tilde{R}_{\mathbf{q}}^a\left(\delta_{ab}-\frac{\tilde{R}_{\mathbf{q}}^b Z_b{}^2 n_b u(\mathbf{q})}{T_b}\right)\end{aligned} \tag{182}$$

where $T_b$ is the temperature associated with component b; and $u(\mathbf{q})=u_0(\mathbf{q})/\varepsilon(\mathbf{q},0)$ is the statically-screened Coulomb potential. Then, using that $w_a=n_aZ_a{}^2T_{\mathrm{e}}/n_eT_a$, which follows from (176),

$$\begin{aligned}S^{(0)}(\mathbf{q})&=\frac{T_{\mathrm{e}}}{n_{\mathrm{e}}}\sum_{a,b}\frac{\tilde{R}_{\mathbf{q}}^a Z_a{}^2 n_a}{T_a}\left(\delta_{ab}-\frac{\tilde{R}_{\mathbf{q}}^b Z_b{}^2 n_b}{T_b}u(\mathbf{q})\right)\\&=\sum_{a,b}S_{ab}^{(0)}(\mathbf{q})\end{aligned} \tag{183}$$

where

$$S_{ab}^{(0)}(\mathbf{q}) = \frac{T_e}{n_e} \frac{\tilde{R}_{\mathbf{q}}^a {Z_a}^2 n_a}{T_a} \left( \delta_{ab} - \frac{\tilde{R}_{\mathbf{q}}^b {Z_b}^2 n_b}{T_b} u(\mathbf{q}) \right)$$
$$= q^2 {D_e}^2 \alpha_a \left( \delta_{ab} - \frac{\alpha_b}{\varepsilon(\mathbf{q},0)} \right) \tag{184}$$

So, for $a = b$,

$$S_{aa}^{(0)}(\mathbf{q}) = \frac{T_e}{n_e} \frac{\tilde{R}_{\mathbf{q}}^a {Z_a}^2 n_a}{T_a} \left( 1 - \tilde{R}_{\mathbf{q}}^a {Z_a}^2 n_a \frac{u(\mathbf{q})}{T_a} \right)$$
$$= w_a \tilde{R}_{\mathbf{q}}^a \left( 1 - \frac{\alpha_{\mathbf{q}}^a}{\varepsilon(\mathbf{q},0)} \right)$$
$$= w_a \tilde{R}_{\mathbf{q}}^a \left( 1 - \frac{\alpha_{\mathbf{q}}^a}{1+\alpha_{\mathbf{q}}} \right) \tag{185}$$
$$= q^2 {D_e}^2 \alpha_{\mathbf{q}}^a \left( 1 - \frac{\alpha_{\mathbf{q}}^a}{1+\alpha_{\mathbf{q}}} \right)$$

which contains a term representing the quantum-statistical correlations between would-be non-interacting identical particles, and a negative Coulomb correlation term.

For $a \neq b$,

$$S_{ab}^{(0)}(\mathbf{q}) = -\frac{T_e}{n_e} \frac{\tilde{R}_{\mathbf{q}}^a {Z_a}^2 n_a}{T_a} \frac{\tilde{R}_{\mathbf{q}}^b {Z_b}^2 n_b}{T_b} u(\mathbf{q})$$
$$= -\frac{q^2 T_e}{m_e {\Omega_e}^2} \frac{\alpha_a \alpha_b}{\varepsilon(\mathbf{q},0)} \tag{186}$$
$$= -q^2 {D_e}^2 \frac{\alpha_a \alpha_b}{1+\alpha}$$

Equation (186) represents the principal correlations between dissimilar particles. The important features of this formula are that it is symmetrical between a and b; and that it vanishes in the absence of any interaction between the different particle species. Note that the Coulomb correlations between like particles (185) and those between unlike particles (186) are both mediated by the statically screened potential, $u(\mathbf{q}) = u_0(\mathbf{q})/\varepsilon(\mathbf{q},0)$. Introducing the elementary 'quasi-classical' static structure factor for species *a*,

$$S_0^a(\mathbf{q}) = 1 - \frac{\alpha_{\mathbf{q}}^a}{1+\alpha_{\mathbf{q}}} \tag{187}$$

which is a form originally proposed by Watson [38], the above become:

$$S_{aa}^{(0)}(\mathbf{q}) = w_a \tilde{R}_{\mathbf{q}}^a S_0^a(\mathbf{q}) = q^2 D_{\mathrm{e}}{}^2 \alpha_{\mathbf{q}}^a S_0^a(\mathbf{q}) \tag{188}$$

$$S_{ab}^{(0)}(\mathbf{q}) = -q^2 D_{\mathrm{e}}{}^2 \varepsilon(\mathbf{q},0)\left(1 - S_0^a(\mathbf{q})\right)\left(1 - S_0^b(\mathbf{q})\right) \ : \ a \neq b \tag{189}$$

A formula that embraces both $a = b$ and $a \neq b$ cases is

$$S_{ab}^{(0)}(\mathbf{q}) = -w_a \tilde{R}_{\mathbf{q}}^a \left(1 - S_0^b(\mathbf{q}) - \delta_{ab}\right) \tag{190}$$

which is implicitly symmetrical under $a \leftrightarrow b$.

It may be helpful to present these equations in a form that makes the *q*-dependence more explicit. Writing

$$\alpha_a = \frac{1}{q^2 \mathfrak{D}_a(q)^2}; \quad \alpha = \frac{1}{q^2 \mathfrak{D}(q)^2} \tag{191}$$

etc, where the $\mathfrak{D}$s are the generalized screening lengths, yields

$$S_{ab}^{(0)}(\mathbf{q}) = \frac{D_{\mathrm{e}}{}^2}{\mathfrak{D}_a{}^2}\left(\delta_{ab} - \frac{1}{q^2 \mathfrak{D}_b{}^2 + \mathfrak{D}_b{}^2/\mathfrak{D}^2}\right) \tag{192}$$

which becomes, in the special case of a two-component plasma (for which $\mathfrak{D}^{-2} = \mathfrak{D}_{\mathrm{i}}{}^{-2} + \mathfrak{D}_{\mathrm{e}}{}^{-2}$)

$$S_{\mathrm{ei}}^{(0)} = -\frac{D_{\mathrm{e}}{}^2}{q^2 \mathfrak{D}_{\mathrm{i}}{}^2 \mathfrak{D}_{\mathrm{e}}{}^2 + \mathfrak{D}_{\mathrm{e}}{}^2 + \mathfrak{D}_{\mathrm{i}}{}^2} = -\tilde{R}_{\mathbf{q}}^{\mathrm{e}} \frac{1}{q^2 \mathfrak{D}_{\mathrm{i}}{}^2 + \mathfrak{D}_{\mathrm{i}}{}^2/\mathfrak{D}^2} = -\tilde{R}_{\mathbf{q}}^{\mathrm{e}}\left(1 - S_0^{\mathrm{i}}(\mathbf{q})\right)$$

$$S_{\mathrm{ee}}^{(0)} = \tilde{R}_{\mathbf{q}}^{\mathrm{e}}\left(1 - \frac{1}{q^2 \mathfrak{D}_{\mathrm{e}}{}^2 + \mathfrak{D}_{\mathrm{e}}{}^2/\mathfrak{D}^2}\right) = \tilde{R}_{\mathbf{q}}^{\mathrm{e}} S_0^{\mathrm{e}}(\mathbf{q}) \tag{193}$$

$$S_{\mathrm{ii}}^{(0)} = \frac{Z^2 n_{\mathrm{i}} T_{\mathrm{e}}}{n_{\mathrm{e}} T_{\mathrm{i}}} \tilde{R}_{\mathbf{q}}^{\mathrm{i}}\left(1 - \frac{1}{q^2 \mathfrak{D}_{\mathrm{i}}{}^2 + \mathfrak{D}_{\mathrm{i}}{}^2/\mathfrak{D}^2}\right) = w_{\mathrm{i}} \tilde{R}_{\mathbf{q}}^{\mathrm{i}} S_0^{\mathrm{i}}(\mathbf{q})$$

which incorporate the generalizations of the classical result for $S_{\mathrm{ee}}^{(0)}$ given by Watson[38].

The radial pair correlation function that follows from (184) in accordance with (175) is

$$g_{\mathbf{r}}\left(a|b\right)=\frac{-Z_b}{Z_a n_a w_b}\frac{D_{\mathrm{e}}{}^2}{D_a{}^2 D_b{}^2}\frac{1}{\left(2\pi\right)^3}\int\frac{D_{\mathbf{q}}{}^2 R_{\mathbf{q}}^a R_{\mathbf{q}}^b}{1+q^2 D_{\mathbf{q}}{}^2}\mathrm{e}^{-\mathrm{i}\mathbf{q}\cdot\mathbf{r}}\mathrm{d}^3\,\mathbf{q} \tag{194}$$

in which $1/D_{\mathbf{q}}{}^2=\sum_a R_{\mathbf{q}}^a/D_a{}^2$. In the degeneracy-corrected Debye-Hückel approximation (in which $\alpha_{\mathbf{q}}^a\simeq R_{\mathbf{0}}^a/q^2 D_a{}^2=\left(1/q\mathfrak{D}_a\left(\mathbf{0}\right)\right)^2$), (194) becomes

$$\begin{aligned}g_{\mathbf{r}}\left(a|b\right)&\simeq\frac{-Z_b R_{\mathbf{0}}^a R_{\mathbf{0}}^b}{Z_a n_a w_b}\frac{D^2 D_{\mathrm{e}}{}^2}{D_a{}^2 D_b{}^2}\frac{1}{\left(2\pi\right)^3}\int\frac{\mathrm{e}^{-\mathrm{i}\mathbf{q}\cdot\mathbf{r}}}{1+q^2 D^2}\mathrm{d}^3\,\mathbf{q}\\&=\frac{-Z_b R_{\mathbf{0}}^a R_{\mathbf{0}}^b}{4\pi Z_a n_a}\frac{\mathrm{e}^{-r/D}}{rD_a{}^2}\\&=-\frac{R_{\mathbf{0}}^a R_{\mathbf{0}}^b}{T_a}\frac{Z_a Z_b e^2}{4\pi\varepsilon_0 r}\mathrm{e}^{-r/D}\end{aligned} \tag{195}$$

in which $1/D^2=\sum_a R_{\mathbf{0}}^a/D_a{}^2$, and where use has been made of (41), (15) and (176). Equations (194) and (195) apply for cases of $a=b$ as well as $a\neq b$.

The function $S^{(1)}\left(\mathbf{q}\right)$ defined by (171) is considered next. This relates to the collective modes associated with each individual species and is written as

$$S^{(1)}\left(\mathbf{q}\right)=\sum_{a,b}S_{ab}^{(1)} \tag{196}$$

where

$$\begin{aligned}S_{ab}^{(1)}\left(\mathbf{q}\right)&=\frac{1}{2}\left\{\lambda_{ab}w_a\tilde{R}_{\mathbf{q}}^a\frac{1-F_a\left(\alpha_a,\beta_a\right)}{\alpha_a G_a}\left(\frac{\Omega_{\mathbf{q}}^a}{2T_a}\coth\frac{\Omega_{\mathbf{q}}^a}{2T_a}-1\right)+\text{transpose}\left(a\leftrightarrow b\right)\right\}\\&=\frac{1}{2}\left\{\lambda_{ab}w_a\tilde{R}_{\mathbf{q}}^a\frac{1}{1+\alpha_a+\beta_a}\frac{\alpha_a}{\alpha_a G_a-1}\left(\frac{\Omega_{\mathbf{q}}^a}{2T_a}\coth\frac{\Omega_{\mathbf{q}}^a}{2T_a}-1\right)+\text{transpose}\left(a\leftrightarrow b\right)\right\}\\&=\frac{1}{2}\left\{\lambda_{ab}w_a\tilde{R}_{\mathbf{q}}^a\left(1-S_0^a\left(\mathbf{q}\right)\right)\frac{1+\alpha}{1+\alpha_a+\beta_a}\frac{1}{\alpha_a G_a-1}\left(\frac{\Omega_{\mathbf{q}}^a}{2T_a}\coth\frac{\Omega_{\mathbf{q}}^a}{2T_a}-1\right)+\text{transpose}\left(a\leftrightarrow b\right)\right\}\end{aligned} \tag{197}$$

where

$$\lambda_a=\sum_b\lambda_{ab} \tag{198}$$

and where $\lambda_{ab} = 0$ for $b \neq a$ if there is no interaction between species $a$ and species $b$, as would generally be the case if $\alpha_a = 0$, or if species $b$ carries no charge, for example. The matrix $S^{(1)}_{ab}$ is defined to be symmetric, although the matrix $\lambda_{ab}$ need not be symmetric.

The third term $S^{(2)}(\mathbf{q})$, defined by (172), is

$$\begin{aligned} S^{(2)}(\mathbf{q}) &\equiv \sum_a w_a \left(R^a_{\mathbf{q}} - \tilde{R}^a_{\mathbf{q}}\right)\left(\frac{1}{1+\alpha} + \left(\lambda_a \frac{1+\beta_a}{1+\alpha_a+\beta_a} - \frac{1}{1+\alpha}\right)\right) \\ &= \sum_{a,b} S^{(2)}_{ab} \end{aligned} \tag{199}$$

which, when combined with (170), yields

$$S^{(0)}(\mathbf{q}) + S^{(2)}(\mathbf{q}) = \sum_a \lambda_a w_a R^a_{\mathbf{q}} \left(\frac{1+\beta_a}{1+\alpha_a+\beta_a}\right) \tag{200}$$

which now contains the correct non-interacting limit. However, it is not amenable to the particle decomposition as applied above to $S^{(0)}(\mathbf{q})$ alone. Noting that $\sum_b \left(S^b_0(\mathbf{q}) - 1 + \delta_{ab}\right) = 1/(1+\alpha)$, we have

$$S^{(2)}_{ab} = \frac{1}{2}\left\{ w_a \left(R^a_{\mathbf{q}} - \tilde{R}^a_{\mathbf{q}}\right)\left(\left(S^b_0(\mathbf{q}) - 1 + \delta_{ab}\right) + \left(1-\delta_{ab}\right)\left(\lambda_{ab}\frac{1+\beta_a}{1+\alpha_a+\beta_a} - \frac{\Delta_{ab}}{1+\alpha}\right)\right) + \text{transpose}(a \leftrightarrow b)\right\} \tag{201}$$

in which $\Delta_{ab}$ is defined to have the properties that, like $\lambda_{ab}$, $\Delta_{ab}$ vanishes when species $a$ and $b$ do not interact; and

$$\sum_{b(\neq a)} \Delta_{ab} = 1 - \lambda_{aa} \frac{(1+\beta_a)(1+\alpha)}{1+\alpha_a+\beta_a} \tag{202}$$

which vanishes, as required by the first property, when $a$ represents an OCP.

The remaining term $S^{(3)}(\mathbf{q})$, defined by (173), can be dealt with similarly to $S^{(1)}$. Using (111) and (187),

$$\begin{aligned} S^{(3)}(\mathbf{q}) &= -\sum_a \lambda_a w_a \left(R^a_{\mathbf{q}} - \tilde{R}^a_{\mathbf{q}}\right) \frac{\alpha_a}{(\alpha_a G_a - 1)(1+\alpha_a+\beta_a)} \\ &= -\sum_a \lambda_a w_a \left(R^a_{\mathbf{q}} - \tilde{R}^a_{\mathbf{q}}\right)\left(1 - S^a_0(\mathbf{q})\right)\left(\frac{1+\alpha}{1+\alpha_a+\beta_a}\right)\left(\frac{1}{\alpha_a G_a - 1}\right) \\ &= \sum_{a,b} S^{(3)}_{ab}(\mathbf{q}) \end{aligned} \tag{203}$$

where

$$S_{ab}^{(3)}(\mathbf{q})=-\frac{1}{2}\left\{\lambda_{ab}w_a\left(R_{\mathbf{q}}^a-\tilde{R}_{\mathbf{q}}^a\right)\left(1-S_0^a(\mathbf{q})\right)\left(\frac{1+\alpha}{1+\alpha_a+\beta_a}\right)\left(\frac{1}{\alpha_a G_a-1}\right)+\text{transpose}(a\leftrightarrow b)\right\} \quad (204)$$

The terms $S^{(2)}(\mathbf{q})$ and $S^{(3)}(\mathbf{q})$ can be treated approximately, by noting that the coefficient $R_{\mathbf{q}}^a-\tilde{R}_{\mathbf{q}}^a$ is $\mathcal{O}\left(q^2/2m_aT\right)$ and hence the summation is dominated by the term corresponding to the lightest particles (assumed to be electrons). Hence,

$$\begin{aligned} S^{(3)}(\mathbf{q}) &\simeq -\left(R_{\mathbf{q}}^{\mathrm{e}}-\tilde{R}_{\mathbf{q}}^{\mathrm{e}}\right)\left(1-S_0^{\mathrm{e}}(\mathbf{q})\right)\left(\frac{1+\alpha}{1+\alpha_{\mathrm{e}}}\right)\left(\frac{1}{\alpha_{\mathrm{e}}G_{\mathrm{e}}-1}\right) \\ &=\sum_{a,b}S_{ab}^{(3)} \end{aligned} \quad (205)$$

where now

$$S_{ab}^{(3)}\simeq-\left(R_{\mathbf{q}}^{\mathrm{e}}-\tilde{R}_{\mathbf{q}}^{\mathrm{e}}\right)\left(1-S_0^{\mathrm{e}}(\mathbf{q})\right)\left(\frac{1+\alpha}{1+\alpha_{\mathrm{e}}}\right)\left(\frac{1}{\alpha_{\mathrm{e}}G_{\mathrm{e}}-1}\right)\delta_{ae}\delta_{be} \quad (206)$$

Similarly,

$$S_{ab}^{(2)}\simeq\left(R_{\mathbf{q}}^{\mathrm{e}}-\tilde{R}_{\mathbf{q}}^{\mathrm{e}}\right)\frac{1}{1+\alpha_{\mathrm{e}}}\delta_{a\mathrm{e}}\delta_{b\mathrm{e}} \quad (207)$$

However, for the present, we avoid making any such approximations.

Finally, collecting up terms,

$$\begin{aligned} S_{ab}(\mathbf{q}) &= S_{ab}^{(0)}(\mathbf{q})+S_{ab}^{(1)}(\mathbf{q})+S_{ab}^{(2)}(\mathbf{q})+S_{ab}^{(3)}(\mathbf{q}) \\ &=\frac{1}{2}\left\{w_aR_{\mathbf{q}}^a\left(S_0^b(\mathbf{q})-1+\delta_{ab}\right)+\text{transpose}(a\leftrightarrow b)\right\} \\ &+\frac{1}{2}\left\{\lambda_{ab}w_a\left(1-S_0^a(\mathbf{q})\right)\frac{1+\alpha}{1+\alpha_a+\beta_a}\frac{1}{\alpha_aG_a-1}\left(\tilde{R}_{\mathbf{q}}^a\frac{\Omega_{\mathbf{q}}^a}{2T_a}\coth\frac{\Omega_{\mathbf{q}}^a}{2T_a}-R_{\mathbf{q}}^a\right)+\text{transpose}(a\leftrightarrow b)\right\} \\ &+\frac{1-\delta_{ab}}{2}\left\{w_a\left(R_{\mathbf{q}}^a-\tilde{R}_{\mathbf{q}}^a\right)\left(\lambda_{ab}\frac{1+\beta_a}{1+\alpha_a+\beta_a}-\frac{\Delta_{ab}}{1+\alpha}\right)+\text{transpose}(a\leftrightarrow b)\right\} \end{aligned} \quad (208)$$

in terms of which,

$$S(\mathbf{q})=\sum_{a,b}S_{ab}(\mathbf{q}) \quad (209)$$

The decomposition (208) has the following essential properties: (1) $S_{aa}(\mathbf{q}) = w_a R_{\mathbf{q}}^a$ if $\alpha_a = 0$; and (2) $S_{ab}(\mathbf{q}) = 0$ if species $b$ is uncharged. Subject to the constraints on the coefficients $\{\lambda\}$, $S_{ab}(\mathbf{q})$ is also finite for all $\mathbf{q} \geq \mathbf{0}$. In particular, it is bounded for any or all of the $\{\alpha, \beta\}$ tending to infinity (ie, $\mathbf{q} \to \mathbf{0}$).

In the case of the two component plasma, for which, referring to (147),

$$
\begin{aligned}
\lambda_{\mathrm{ee}} &= 1 \\
\lambda_{\mathrm{ii}} &= \left(\frac{1}{1+\alpha_{\mathrm{e}}}\right)^2 \\
\lambda_{\mathrm{ei}} &= \left(\frac{\Omega_{\mathrm{i}}}{\Omega_{\mathrm{e}}}\right)^2 \left(1 - \left(\frac{1}{1+\alpha_{\mathrm{e}}}\right)^2\right) \\
\lambda_{\mathrm{ie}} &= 0
\end{aligned}
\tag{210}
$$

together with $\beta_{\mathrm{e}} = 0$, $\beta_{\mathrm{i}} = \alpha_{\mathrm{e}}$, the coefficients $\Delta_{ab}$ can be straightforwardly deduced from (202) to be as follows

$$
\begin{aligned}
\Delta_{\mathrm{ei}} &= -\frac{\alpha_{\mathrm{i}}}{1+\alpha_{\mathrm{e}}} \\
\Delta_{\mathrm{ie}} &= \frac{\alpha_{\mathrm{e}}}{1+\alpha_{\mathrm{e}}}
\end{aligned}
\tag{211}
$$

and hence

$$
S_{\mathrm{ee}}(\mathbf{q}) = R_{\mathbf{q}}^{\mathrm{e}} S_0^{\mathrm{e}}(\mathbf{q}) + \left(1 - S_0^{\mathrm{e}}(\mathbf{q})\right)\left(\frac{1+\alpha}{1+\alpha_{\mathrm{e}}}\right)\frac{1}{\alpha_{\mathrm{e}} G_{\mathrm{e}} - 1}\left(\tilde{R}_{\mathbf{q}}^{\mathrm{e}} \frac{\Omega_{\mathbf{q}}^{\mathrm{e}}}{2T_{\mathrm{e}}} \coth \frac{\Omega_{\mathbf{q}}^{\mathrm{e}}}{2T_{\mathrm{e}}} - R_{\mathbf{q}}^{\mathrm{e}}\right) \tag{212}
$$

$$
S_{\mathrm{ii}}(\mathbf{q}) = w_{\mathrm{i}}\left(R_{\mathbf{q}}^{\mathrm{i}} S_0^{\mathrm{i}}(\mathbf{q}) + \left(1 - S_0^{\mathrm{i}}(\mathbf{q})\right)\left(\frac{1}{1+\alpha_{\mathrm{e}}}\right)^2 \frac{1}{\alpha_{\mathrm{i}} G_{\mathrm{i}} - 1}\left(\tilde{R}_{\mathbf{q}}^{\mathrm{i}} \frac{\Omega_{\mathbf{q}}^{\mathrm{i}}}{2T_{\mathrm{i}}} \coth \frac{\Omega_{\mathbf{q}}^{\mathrm{i}}}{2T_{\mathrm{i}}} - R_{\mathbf{q}}^{\mathrm{i}}\right)\right) \tag{213}
$$

$$
\begin{aligned}
S_{\mathrm{ei}}(\mathbf{q}) = S_{\mathrm{ie}}(\mathbf{q}) = &\frac{1}{2}\left\{R_{\mathbf{q}}^{\mathrm{e}}\left(S_0^{\mathrm{i}}(\mathbf{q}) - 1\right) + w_{\mathrm{i}} R_{\mathbf{q}}^{\mathrm{i}}\left(S_0^{\mathrm{e}}(\mathbf{q}) - 1\right)\right\} \\
&+ \frac{\lambda_{\mathrm{ei}}}{2(1+\alpha_{\mathrm{e}})}\left\{\left(R_{\mathbf{q}}^{\mathrm{e}} - \tilde{R}_{\mathbf{q}}^{\mathrm{e}}\right) + \left(1 - S_0^{\mathrm{e}}(\mathbf{q})\right)\frac{1+\alpha}{\alpha_{\mathrm{e}} G_{\mathrm{e}} - 1}\left(\tilde{R}_{\mathbf{q}}^{\mathrm{e}} \frac{\Omega_{\mathbf{q}}^{\mathrm{e}}}{2T_a} \coth \frac{\Omega_{\mathbf{q}}^{\mathrm{e}}}{2T_a} - R_{\mathbf{q}}^{\mathrm{e}}\right)\right\} \\
&+ \frac{1}{2(1+\alpha)(1+\alpha_{\mathrm{e}})}\left\{\alpha_{\mathrm{i}}\left(R_{\mathbf{q}}^{\mathrm{e}} - \tilde{R}_{\mathbf{q}}^{\mathrm{e}}\right) - w_{\mathrm{i}} \alpha_{\mathrm{e}}\left(R_{\mathbf{q}}^{\mathrm{i}} - \tilde{R}_{\mathbf{q}}^{\mathrm{i}}\right)\right\}
\end{aligned}
\tag{214}
$$

in which $S_0^a(\mathbf{q})$ is the quasi-classical structure factor for species $a$ as given by (187), $\lambda_{\mathrm{ei}}$ is given by (210) and where, from (118),

$$\alpha_{\mathrm{i}}G_{\mathrm{i}}-1=\frac{\alpha_{\mathrm{i}}}{1+\alpha_{\mathrm{e}}}+\left(\alpha_{\mathrm{i}}G_{\mathrm{i}}^{0}-1-\alpha_{\mathrm{i}}\right)\left(1+\frac{\alpha_{\mathrm{i}}\alpha_{\mathrm{e}}}{\left(\alpha_{\mathrm{i}}G_{\mathrm{i}}^{0}-1-\alpha_{\mathrm{i}}\right)+\left(1+\alpha_{\mathrm{i}}\right)\left(\alpha_{\mathrm{i}}G_{\mathrm{i}}^{0}-1\right)}\right)$$

$$\alpha_{\mathrm{e}}G_{\mathrm{e}}-1=\alpha_{\mathrm{e}}G_{\mathrm{e}}^{0}-1 \tag{215}$$

where $G_a=\left(\Omega_{\mathrm{q}}^{a}/\Omega_a\right)^2$ and $G_a^0$ is the value of this quantity that is appropriate for a one component system comprising particle type $a$.

The final term in (214) can be further simplified, using (177), to yield, without further approximation, that

$$\alpha_{\mathrm{i}}\left(R_{\mathbf{q}}^{\mathrm{e}}-\tilde{R}_{\mathbf{q}}^{\mathrm{e}}\right)-w_{\mathbf{i}}\alpha_{\mathrm{e}}\left(R_{\mathbf{q}}^{\mathrm{i}}-\tilde{R}_{\mathbf{q}}^{\mathrm{i}}\right)=\alpha_{\mathrm{i}}\left(R_{\mathbf{q}}^{\mathrm{e}}-\frac{\tilde{R}_{\mathbf{q}}^{\mathrm{e}}R_{\mathbf{q}}^{\mathrm{i}}}{\tilde{R}_{\mathbf{q}}^{\mathrm{i}}}\right) \tag{216}$$

Moreover the ions in a plasma are typically assumed to obey a Boltzmann distribution, in which case $R_{\mathbf{q}}^{\mathrm{i}}=1$, while it is generally reasonable to neglect the term proportional to $\lambda_{\mathrm{ei}}$, since it is $\mathcal{O}\left(m_{\mathrm{e}}/m_{\mathrm{i}}\right)$. Hence, making these substitutions and approximations, equations (213) - (214) become

$$S_{\mathrm{ii}}(\mathbf{q})=w_{\mathrm{i}}\left(S_0^{\mathrm{i}}(\mathbf{q})+\left(1-S_0^{\mathrm{i}}(\mathbf{q})\right)\left(\frac{1}{1+\alpha_{\mathrm{e}}}\right)^2\frac{1}{\alpha_{\mathrm{i}}G_{\mathrm{i}}-1}\left(\tilde{R}_{\mathbf{q}}^{\mathrm{i}}\frac{\Omega_{\mathbf{q}}^{\mathrm{i}}}{2T_{\mathrm{i}}}\coth\frac{\Omega_{\mathbf{q}}^{\mathrm{i}}}{2T_{\mathrm{i}}}-1\right)\right) \tag{217}$$

$$S_{\mathrm{ei}}(\mathbf{q})=S_{\mathrm{ie}}(\mathbf{q})=\frac{1}{2}\left\{R_{\mathbf{q}}^{\mathrm{e}}\left(S_0^{\mathrm{i}}(\mathbf{q})-1\right)+w_{\mathrm{i}}\left(S_0^{\mathrm{e}}(\mathbf{q})-1\right)+\frac{\alpha_{\mathrm{i}}}{(1+\alpha)(1+\alpha_{\mathrm{e}})}\left(R_{\mathbf{q}}^{\mathrm{e}}-\tilde{R}_{\mathbf{q}}^{\mathrm{e}}/\tilde{R}_{\mathbf{q}}^{\mathrm{i}}\right)\right\} \tag{218}$$

These equations are valid for arbitrary values of $\mathbf{q}$. In the regime of ‘small’ $\mathbf{q}$, the following approximate forms are considered applicable:

$$S_{\mathrm{ee}}(\mathbf{q})\simeq R_{\mathbf{q}}^{\mathrm{e}}S_0^{\mathrm{e}}(\mathbf{q})+R_{\mathbf{0}}^{\mathrm{e}}\left(1-S_0^{\mathrm{e}}(\mathbf{q})\right)\left(\frac{1+\alpha}{1+\alpha_{\mathrm{e}}}\right)\frac{1}{\alpha_{\mathrm{e}}G_{\mathrm{e}}-1}\left(\frac{\Omega_{\mathbf{q}}^{\mathrm{e}}}{2T_{\mathrm{e}}}\coth\frac{\Omega_{\mathbf{q}}^{\mathrm{e}}}{2T_{\mathrm{e}}}-1\right) \tag{219}$$

$$S_{\mathrm{ii}}(\mathbf{q})\simeq w_{\mathrm{i}}\left(S_0^{\mathrm{i}}(\mathbf{q})+\left(1-S_0^{\mathrm{i}}(\mathbf{q})\right)\left(\frac{1}{1+\alpha_{\mathrm{e}}}\right)^2\frac{1}{\alpha_{\mathrm{i}}G_{\mathrm{i}}-1}\left(\frac{\Omega_{\mathbf{q}}^{\mathrm{i}}}{2T_{\mathrm{i}}}\coth\frac{\Omega_{\mathbf{q}}^{\mathrm{i}}}{2T_{\mathrm{i}}}-1\right)\right) \tag{220}$$

$$S_{\mathrm{ei}}(\mathbf{q}) = S_{\mathrm{ie}}(\mathbf{q}) = \frac{1}{2}\left\{ R_{\mathbf{q}}^{\mathrm{e}}\left(S_0^{\mathrm{i}}(\mathbf{q})-1\right) + w_{\mathrm{i}}\left(S_0^{\mathrm{e}}(\mathbf{q})-1\right) + \frac{\alpha_{\mathrm{i}}}{(1+\alpha)(1+\alpha_{\mathrm{e}})}\left(R_{\mathbf{q}}^{\mathrm{e}} - \tilde{R}_{\mathbf{q}}^{\mathrm{e}}\right)\right\} \tag{221}$$

in which quantal terms up to at least $\mathcal{O}\left(q^2/m_{\mathrm{e}}T_{\mathrm{e}}\right)$ have been retained. An appropriate approximation, in this context, for $R_{\mathbf{q}}^{\mathrm{e}} - \tilde{R}_{\mathbf{q}}^{\mathrm{e}}$ is given by (47).

Equations (208)-(221) comprise the quasi-classical – quantal decompositions of the static structure factor. It is of the form, for example,

$$S_{aa}(\mathbf{q}) = w_a\tilde{R}_a\left( S_0^a + \left(1-S_0^a\right)Y_a\left(\frac{\Omega_{\mathbf{q}}^a}{2T_a}\coth\frac{\Omega_{\mathbf{q}}^a}{2T_a} - 1\right)\right) + \mathcal{O}\left(R_a - \tilde{R}_a\right)$$

and should not be confused with the modal decomposition (168) – which has the shape

$$S^a(\mathbf{q}) = \lambda_a w_a\left( R_aF_a + \tilde{R}_a\left(1-F_a\right)X_a\frac{\Omega_{\mathbf{q}}^a}{2T_a}\coth\frac{\Omega_{\mathbf{q}}^a}{2T_a}\right)$$

#### 4.2.5 Effective Collision Frequency

A quantity not so far discussed is the effective collision frequency, or damping frequency, defined at equation (104). This determines the location of the zeros (resonances) of the dielectric function in the lower half of the complex frequency plane. In order to determine the properties of this parameter, and the physical basis for it, we need to examine the properties of the dielectric function in the region of a resonance. From equations (101) and (105), it can be seen that, close to a resonance at $z = G_a$,

$$-\frac{(1+\beta)^2}{\varepsilon_a + \beta} \sim \frac{1 - F_a(\alpha, \beta)}{G_a - z} \tag{222}$$

while taking the imaginary part of (105) close to the resonance yields

$$\frac{(1+\beta)^2}{|\varepsilon_a + \beta|^2} \operatorname{Im} \varepsilon_a \sim \frac{1 - F_a}{|G_a - z|^2} \operatorname{Im} z \tag{223}$$

Hence, upon combining (222) and (223) and evaluating the expressions at $\Omega^a_{\alpha\beta} = G_a(\alpha_a, \beta_a)\Omega_a$, we have

$$\frac{\nu^a_{\alpha\beta}}{\Omega_a} = \frac{\Omega_a}{\Omega^a_{\alpha\beta}} \operatorname{Im} z \simeq \frac{\Omega_a}{\Omega^a_{\alpha\beta}} \frac{1 - F_a}{(1+\beta_a)^2} \operatorname{Im} \varepsilon_a\left(\alpha, \Omega^a_{\alpha\beta}\right) \tag{224}$$

which is an important relation connecting the damping frequency to the imaginary part of the dielectric function at a real frequency in the neighbourhood of the resonance. In the SPPA model of an OCP ($F_a = 0$, $\beta_a = 0$) equation (224) reduces to $\nu^a_{\alpha\beta} \simeq \Omega_a \operatorname{Im} \varepsilon_a\left(\alpha, \Omega^a_{\alpha\beta}\right) / G_a(\alpha_a, \beta_a)$. Equation (224) does however incorporate the essential generalizations for treating multicomponent plasmas in the GPPA, as discussed in section 4.2.

The following subsection evaluates this expression in the RPA to derive explicit formulae for the Landau damping frequency.

### 4.2.6 The Effective Collision Frequency in the RPA – Landau Damping

Equation (224) can be evaluated using the RPA. Since the RPA describes a plasma in which short range collisions do not occur, the resulting effective collision frequency describes only Landau damping, which is a manifestation of mode mixing between particle translational motion and the long-wavelength collective modes. Using (55), this yields

$$\frac{\nu^a_{\alpha\beta}}{\Omega_a} \simeq \frac{\pi}{8}\frac{1-F_a}{\left(1+\beta_a\right)^2}\left(\frac{\Omega_a}{T}\right)^3 \frac{1}{\nu^{3/2}}\frac{1}{I_{1/2}\left(\eta_a\right)}\frac{L\left(u^a_{\alpha\beta},\nu_a;\eta_a\right)}{u^a_{\alpha\beta}}$$
$$= \frac{\pi}{2\sqrt{2}}\frac{1-F_a}{\left(1+\beta_a\right)^2}\frac{1}{\left(qD_a\right)^3}\frac{1}{I_{1/2}\left(\eta_a\right)}\frac{L\left(u^a_{\alpha\beta},\nu_a;\eta_a\right)}{u^a_{\alpha\beta}} \tag{225}$$

where $u^a_{\alpha\beta} = \Omega^a_{\alpha\beta}/T$ , $\nu = q^2/2m_aT$ and $\eta_a$ is the degeneracy parameter for species *a*. Applying the approximation (58) yields

$$\frac{\nu^a_{\alpha\beta}}{\Omega_a} \simeq \frac{\pi}{2\sqrt{2}}\frac{1-F_a}{\left(1+\beta_a\right)^2}\frac{1}{\left(qD_a\right)^3}\frac{\sinh\left(\frac{1}{2}u^a_{\alpha\beta}\right)}{\frac{1}{2}u^a_{\alpha\beta}}\frac{1}{I_{1/2}\left(\eta_a\right)}\left(1+\exp\left(\frac{q^2}{8m_aT}+\frac{G_a}{2\left(qD_a\right)^2}-\eta_a\right)\right)^{-1} \tag{226}$$

This function vanishes in both limits, $q \to 0$ and $q \to \infty$. (Note that the quantal term, $q^2/8m_aT$, in the exponent becomes important in controlling this function at large *q*.) Through the presence of the parameter $\beta_a$, equation (226) provides the Landau damping frequency of component $a$ in a multicomponent plasma (as described in section 4.2) when $F_a$ and $G_a$ are given by (111) and (118) respectively. The (complex) plasma resonance frequency is then found to be

$$\hat{\Omega}^a_{\alpha\beta} = \sqrt{\left(\Omega^a_{\alpha\beta}\right)^2 - \frac{1}{4}\left(\nu^a_{\alpha\beta}\right)^2} - \frac{1}{2}\mathrm{i}\nu^a_{\alpha\beta} = \sqrt{G_a\Omega_a{}^2 - \frac{1}{4}\left(\nu^a_{\alpha\beta}\right)^2} - \frac{1}{2}\mathrm{i}\nu^a_{\alpha\beta} \tag{227}$$

In particular, for a Boltzmann distribution ($\eta << -1$), the formula (226) becomes

$$\frac{\nu^a_{\alpha\beta}}{\Omega_a} \simeq \sqrt{\frac{\pi}{2}}\frac{1-F_a}{\left(1+\beta\right)^2}\frac{1}{\left(qD_a\right)^3}\frac{\sinh\left(\frac{1}{2}u^a_{\alpha\beta}\right)}{\frac{1}{2}u^a_{\alpha\beta}}\exp\left(-\frac{q^2}{8m_aT}-\frac{G_a}{2\left(qD_a\right)^2}\right) \tag{228}$$

In the special case of an OCP ($\beta_a = 0$), $G_a = G^0_a = 1+3\left(qD_a\right)^2$, in accordance with (80), so that

$$\frac{\nu^a_{\alpha 0}}{\Omega_a} \simeq \sqrt{\frac{\pi}{2}}\frac{1-F^0_a}{\left(qD_a\right)^3}\frac{\sinh\left(\frac{1}{2}u^a_{\alpha 0}\right)}{\frac{1}{2}u^a_{\alpha 0}}\exp\left(-\frac{q^2}{8m_aT}-\frac{1}{2\left(qD_a\right)^2}-\frac{3}{2}\right) \tag{229}$$

in which $u^a_{\alpha 0} = \left(\Omega_a/T\right)\left(1+3q^2D_a{}^2\right)^{1/2}$ and $F^0_a$ is given by (114). In the classical limit of an OCP, $\Omega_a/T_a << 1$ and $q^2 << m_aT$, and with $F^0_a = 0$ (SPPA), the result (229) agrees exactly with that given by (91). However equation (229) provides extension to arbitrarily large values of $q$ , both through the quantal term in the exponent, and the factor of $1-F_a^0$ , which, for small $q$, is $1+\mathcal{O}\left(q^2 D_a^{\ 2}\right)$. The damping frequency (229) has a maximum with respect to wavenumber near

$$q^2 D_a^{\ 2} = X \equiv 6\left(T/\Omega_a\right)^2\left(\left(1+\left(\Omega_a/3T\right)^2\right)^{1/2}-1\right) \tag{230}$$

which $\rightarrow 1/3$ at high temperatures when $T >> \Omega_a$ , and $\rightarrow 2T/\Omega_a$ at low temperatures when $T << \Omega_a$. In general therefore, the maximum damping frequency given by (229) is found to be given approximately by

$$\begin{aligned}\left(\frac{\nu_{\alpha 0}^a}{\Omega_a}\right)_{\max} &\approx \sqrt{\frac{\pi}{2}}\left(1-F_a^0\left(X\right)\right)\left(\frac{\Omega_a}{T}\right)^3\left(6\left(\left(1+\left(\Omega_a/3T\right)^2\right)^{1/2}-1\right)\right)^{-3/2} \\ &\times\frac{\sinh\left(\left(1+3X\right)^{1/2}\left(\Omega_a/2T\right)\right)}{\left(1+3X\right)^{1/2}\left(\Omega_a/2T\right)}\exp\left(-\frac{3}{2}\left(\left(1+\left(\Omega_a/3T\right)^2\right)^{1/2}+1\right)\right)\end{aligned} \tag{231}$$

where $F_a^0\left(X\right)=F_a\left(X^{-1},0\right)$, so, in the former case ($T >> \Omega_a,\ X \sim 1/3,\ F_a^0\left(\tfrac{1}{3}\right)=1/10$), the maximum damping frequency is

$$\left(\frac{\nu_{\alpha 0}^a}{\Omega_a}\right)_{\max,\ T>>\Omega} \approx 2.7\sqrt{\frac{3\pi}{2}}\,\mathrm{e}^{-3} = 0.2918... \tag{232}$$

which is a numerical constant. In the other limit ($T << \Omega_a,\ X \sim 2T/\Omega_a << 1,\ F_a^0\left(X\right)\sim 0$),

$$\left(\frac{\nu_{\alpha 0}^a}{\Omega_a}\right)_{\max,\ T<<\Omega} \approx \frac{\sqrt{\pi}}{4}\left(\frac{\Omega_a}{T}\right)^{1/2} \tag{233}$$

which predicts that, in a low temperature non-degenerate quantum plasma, the Landau damping should become large on wavelength scales $>> D_a$, so any collective modes would have very short lifetimes. However, in such a regime, $\mathrm{e}^{-\Omega/T} \ll 1$ , so the equilibrium excitation of collective modes is negligible anyway.

## 5 SUMMARY AND CONCLUSIONS

This article begins by presenting a logically set-out review of the dielectric function of a Coulomb plasma in the context of a modern Green function formalism, within a consistent notation. Two particular approximations are visited: the RPA, which describes weakly coupled systems, and the plasmon-pole approximation, which, in its standard form, describes strongly coupled systems entirely in terms of their resonant modes. The inconsistencies between these two models, in terms of their limits, and the Bohm-Gross relation, are highlighted, showing that they could not be used to describe the transition of a plasma from a strongly coupled regime to a weakly coupled one, for example. A number of new results are presented along the way. These include generalized forms of the f- and screening sum rules, equations (33) and (34). A study of the properties of the key functions $\tilde{R}_{\mathbf{q}}$ and $R_{\mathbf{q}}$ defined by equations (42) and (43), which includes a universal Padé approximant for $\tilde{R}_{\mathbf{q}}$ in the form of equation (73), and their relationship to the static structure factor $S(\mathbf{q})$ and the static dielectric function is presented. The formula (55), incorporating the function (57), for the dielectric function and structure factor of the finite-temperature ideal fermion gas, while not new, is not, it seems, well known. The more tractable approximation to (57)as given by equation (58) is presented as being applicable all regimes except for that of extreme degeneracy, $T \ll T_{\mathrm{F}}$ .

This introductory review provides the basis for a novel analytical model for the dielectric function of multi-component plasmas applicable to ideal and non-ideal regimes, which is then described. The generalized plasmon pole approximation incorporates features of the standard plasmon pole model, but is able to incorporate the RPA limit. The model is capable of considerable flexibility, as many of the spatial properties, which are expressed as parameter dependencies on wave number, are not prescribed, except, of course, in the RPA limit.

The model describes plasmas in which the dynamical strength is shared between independent-particle-like modes and oscillatory collective modes. This modal behaviour is characteristic of Coulomb systems and is a feature of the dielectric function (138)-(139) and of the charge-density structure factor (150). It is also a feature of the RPA, where the mixing of collective plasma modes with free particle modes is responsible for Landau damping. However the GPPA model provides a generalization of the RPA and can describe regimes of much stronger coupling where more of the strength is transferred to collective modes and may therefore be useful for modelling liquid-like or solid-like plasma states.

The model is shown to possess all the analytic properties required of a physical dielectric function satisfying causality and the various sum rules, which relate to behaviour at zero frequency and in the high-frequency limit. Additional singularities in the form of branch points are introduced by the model, but these are confined to the negative imaginary frequency domain and so are consistent with causality. However, the GPPA model per se leaves the spatial properties of the dielectric function largely unspecified, and, for illustration, we have appealed to the RPA for a description of these.

One of the features of the model is its applicability to multicomponent plasmas, most particularly the two component plasma consisting of electrons and ions, in which the particles of one component are much lighter than those of the other. The shift in the resonant frequencies, through coupling due to screening, between the components, is described, in the general multicomponent case, by equations (118) combined with (227), and, in the case of a two-component plasma, by (215). In the RPA limit, these provide a description of ion acoustic modes, which, in the solid state, would translate to phonon modes.

Mode damping is associated with the imaginary part of the resonant frequency. In sections 4.2.5 - 4.2.6 new formulae are presented for the Landau damping frequency, which extend to the quantum regime as well as to partially degenerate plasmas, while including the generalizations to multicomponent plasmas.

The modelled dielectric function is used to derive formulae for the dynamic and static structure factors and both modal and classical-quantal decompositions are derived.

The new model also has implications for the treatment of Thomson scattering [39], [7], and the simulation and interpretation of experiments [40], through the ability to describe Brillouin scattering, which is inelastic scattering of photons due to coupling to the long-wavelength resonant modes of the scattering system. Because the Thomson scattering amplitude is inversely proportional to the mass of the scattering charge, it effectively picks out only the electron component of the structure factor, in the first instance. This provides a description of incoherent scattering [6], [7] by free electrons, which would apply to a fully ionised plasma, for example. This electron component is given by (212), or approximately by (219). However electron-ion correlations mean that multi-atom coherent scattering can also occur [7]. This depends on the ion-ion part of the structure factor, which is given by equation (213) or (220).

The principal result of this report is the generalized plasmon pole approximation as expressed by (138)-(139) in which the functions $F(\alpha,\beta)$ and $G(\alpha,\beta)$ are given by (111)-(112) and (118).

These equations yield the dynamic structure factor according to (150) which is expressed by (163)-(166) and yield the static structure factor given by (168) or (208)-(209). In the case of a two-component plasma consisting of electrons and ions, the static structure factor is provided by (212)-(221).

The model is capable of providing a reasonable description of the RPA regime (through the parameterisations of $\Omega_{\mathbf{q}}, \nu_{\mathbf{q}}, R_{\mathbf{q}}, \tilde{R}_{\mathbf{q}}$, as represented by equations (80), (229), (65) and (73) respectively) with only minor differences of approximation, while, at the same time, being extendable to regimes of stronger coupling. Moreover the model includes the generalization to multi-component plasmas where the shift in the resonance frequencies, eg as provided by (118) and (226), is taken into account through the $\beta$ parameters. These parameters are given, in the electron-ion TCP, by equations (147), which yield the modified resonance frequencies according to (167). These describe the usual Langmuir and ion-acoustic modes, albeit in a somewhat generalized form that takes account of the wavelength dependencies of the underlying single-particle modes.

# APPENDIX A: SUM RULES

## A.1 F-sum rule formula

Let $F(z)$ be a function of a complex variable $z$ that is regular in the domain $\operatorname{Im} z \geq 0$, and for which, in $\operatorname{Im} z > 0$:

$$\lim_{|z|\to\infty}\left[\operatorname{Im} F(z)\right] \equiv \operatorname{Im} F(\infty) = 0 \tag{234}$$

$$\lim_{|z|\to\infty}\left[\operatorname{Im}\left(z^2\left(F(z)-F(\infty)\right)\right)\right] = 0 \tag{235}$$

By Cauchy's integral theorem, the contour integral, $\oint_C z\left(F(z)-F(\infty)\right)\mathrm{d}z$, taken around any contour $C$ in the upper half plane, vanishes. Now let $C$ be the contour taken, along the real axis from $z=-R$, to $z=+R$, returning along the semicircular arc $|z|=R$ in $\pi \geq \arg z \geq 0$. Then

$$\oint_C \left(F(z)-F(\infty)\right) z\,\mathrm{d}z = \int_{-R}^{+R}\left(F(z)-F(\infty)\right) z\,\mathrm{d}z + \mathrm{i}\,R^2\int_0^{\pi}\left(F\left(R\mathrm{e}^{\mathrm{i}\theta}\right)-F(\infty)\right)\mathrm{e}^{2\mathrm{i}\theta}\mathrm{d}\theta = 0 \tag{236}$$

Hence, in the limit $R\to\infty$, when $C$ encloses the whole upper half plane,

$$\int_{-\infty}^{+\infty}\left(F(z)-F(\infty)\right) z\,\mathrm{d}z = -\mathrm{i}\,\pi \lim_{|z|\to\infty}\left[z^2\left(F(z)-F(\infty)\right)\right] \tag{237}$$

Finally, taking the imaginary part, using (234) and (235),

$$\int_{-\infty}^{\infty}\operatorname{Im} F(z)\, z\,\mathrm{d}z = -\pi \lim_{|z|\to\infty}\left[z^2\left(F(z)-F(\infty)\right)\right] \tag{238}$$

Equation (238) is the basic mathematical form of the f-sum rule formula.

## A.2 Screening/compressibility sum rule formula

Let $F(z)$ be a function of a complex variable $z$ that is both regular in the domain $\operatorname{Im} z \geq 0$, and satisfies:

$$\lim_{|z|\to\infty}\left[\operatorname{Im} F(z)\right] \equiv \operatorname{Im} F(\infty) = 0 \quad : \quad 0 < \arg z < \pi \tag{239}$$

$$\operatorname{Im} F(0) = 0 \tag{240}$$

Since $F(z)$ is regular in $\operatorname{Im} z \geq 0$, it follows that $\frac{1}{z}\left(F(z) - F(0)\right)$ is also regular in $\operatorname{Im} z \geq 0$, in which case, by Cauchy's integral theorem,

$$\oint_C \left(F(z) - F(0)\right)\frac{\mathrm{d}z}{z} = 0 \tag{241}$$

for any contour $\mathcal{C}$ in the upper half plane. Taking $\mathcal{C}$ to be the contour as defined above, yields, in the limit $R \to \infty$,

$$\oint_C \left(F(z) - F(0)\right)\frac{\mathrm{d}z}{z} = \int_{-\infty}^{\infty}\left(F(z) - F(0)\right)\frac{\mathrm{d}z}{z} + \mathrm{i}\int_0^{\pi}\left(F(\infty) - F(0)\right)\mathrm{d}\theta = 0 \tag{242}$$

and therefore

$$\int_{-\infty}^{\infty}\left(F(z) - F(0)\right)\frac{\mathrm{d}z}{z} = -\mathrm{i}\pi\left(F(\infty) - F(0)\right) \tag{243}$$

Finally, taking the imaginary part, using (239) and (240),

$$\int_{-\infty}^{\infty} \operatorname{Im} F(z)\frac{\mathrm{d}z}{z} = \pi\left(F(0) - F(\infty)\right) \tag{244}$$

Equation (244) is the basic mathematical form of the screening/compressibility sum-rule formula. If it is further given that $\operatorname{Im} F(z)$ is an odd function of $z$ on $\operatorname{Im} z = 0$, then equations (238) and (244) become

$$\int_0^{\infty} \operatorname{Im} F(z)\, z\, \mathrm{d}z = -\frac{\pi}{2}\lim_{|z|\to\infty}\left[z^2\left(F(z) - F(\infty)\right)\right] \tag{245}$$

$$\int_0^{\infty} \operatorname{Im} F(z)\frac{\mathrm{d}z}{z} = \frac{\pi}{2}\left(F(0) - F(\infty)\right) \tag{246}$$

# APPENDIX B: MATHEMATICAL FORMULAE

## B.1 Proof of Equation (70)

$$\lim_{\eta\to-\infty}\frac{1}{2v^{1/2}}\frac{1}{I_{1/2}(\eta)}\int_0^\infty L(u,v;\eta)\frac{\mathrm{d}u}{u}=\frac{2}{\sqrt{v}}\Phi\left(\frac{\sqrt{v}}{2}\right) \tag{247}$$

where $\Phi(x)$ is Dawson's integral (71).

[Proof:

Let the integral (247) be denoted $R(v)$. Then, using that, for $\eta\to-\infty$

$$I_{1/2}(\eta)\sim\frac{\sqrt{\pi}}{2}\exp(\eta) \tag{248}$$

and

$$\begin{aligned}L(u,v;\eta)&\sim\left(1-\mathrm{e}^{-u}\right)\exp\left(\eta-\frac{(u-v)^2}{4v}\right)\\&=2\exp\left(\eta-\tfrac{1}{4}v\right)\exp\left(-\frac{u^2}{4v}\right)\sinh\left(\frac{u}{2}\right)\end{aligned} \tag{249}$$

we have

$$\begin{aligned}R(v)&\sim\frac{2\mathrm{e}^{-\frac{1}{4}v}}{\sqrt{\pi v}}\int_0^\infty\exp\left(-\frac{u^2}{4v}\right)\sinh\left(\frac{u}{2}\right)\frac{\mathrm{d}u}{u}\\&=\frac{2\mathrm{e}^{-\frac{1}{4}v}}{\sqrt{\pi v}}J\left(\sqrt{v}\right)\end{aligned} \tag{250}$$

where

$$J(x)=\int_0^\infty\exp\left(-\frac{t^2}{x^2}\right)\frac{\sinh(t)}{t}\mathrm{d}t \tag{251}$$

Expanding the hyperbolic sine function as a power series in *t*, and carrying out the integration term by term yields,

$$J(x) = \sum_{n=0}^{\infty} \frac{x^{2n+1}}{(2n+1)!} \int_0^{\infty} \mathrm{e}^{-y^2} y^{2n} \,\mathrm{d}y$$

$$= \frac{1}{2} \sum_{n=0}^{\infty} \frac{x^{2n+1}}{(2n+1)!} \Gamma\left(n+\tfrac{1}{2}\right) \tag{252}$$

$$= \frac{1}{2} \sum_{n=0}^{\infty} \frac{\Gamma\left(n+\frac{1}{2}\right)}{\Gamma(2n+2)} x^{2n+1}$$

where $\Gamma(z)$ denotes the Euler gamma function [41], a standard property of which is

$$\Gamma(2n+2) = (2n+1)2n\Gamma(2n) = \frac{2n(2n+1)}{\sqrt{2\pi}} 2^{2n-1/2} \Gamma(n) \Gamma\left(n+\tfrac{1}{2}\right) \tag{253}$$

Hence

$$J(x) = \sqrt{\pi} \sum_{n=0}^{\infty} \frac{1}{(2n+1)n!} \left(\frac{x}{2}\right)^{2n+1}$$

$$= \sqrt{\pi} \int_0^{x/2} \exp\left(t^2\right) \mathrm{d}t \tag{254}$$

$$= \sqrt{\pi} \exp\left(\tfrac{1}{4} x^2\right) \Phi\left(\tfrac{1}{2} x\right)$$

where $\Phi(x)$ is Dawson's integral (71). Hence, combining (254) and (250)

$$R(v) = \frac{2}{\sqrt{v}} \Phi\left(\frac{\sqrt{v}}{2}\right) \tag{255}$$

QED]

## B.2 Determination of the integral $\int F(\omega)\mathrm{Im}\{1/(\omega+z^*) - 1/(\omega-z)\}\mathrm{d}\omega$

The integral $\int_{-\infty}^{+\infty} F(\omega)\,\mathrm{Im}\left(\frac{1}{\omega+z^*}-\frac{1}{\omega-z}\right)\mathrm{d}\omega$ is here considered, in which $z$ is a complex constant such that $\mathrm{Im}\,z<0$, and $F(\omega)$ is a meromorphic function that is real and finite everywhere on the real $\omega$ axis, except for the possibility of having a pole at $\omega=0$, in which case

$$\int_{-\infty}^{+\infty} F(\omega)\,\mathrm{Im}\left(\frac{1}{\omega+z^*}-\frac{1}{\omega-z}\right)\mathrm{d}\omega=\mathrm{Im}\,\wp\int_{-\infty}^{\infty} F(\omega)\left(\frac{1}{\omega+z^*}-\frac{1}{\omega-z}\right)\mathrm{d}\omega \tag{256}$$

where $\wp$ denotes the principal value at $\omega=0$. To evaluate this integral, we first define the contour $\mathcal{C}_P$ running from $\omega=-\infty$ to $\omega=+\infty$ such that $\mathrm{Im}\left(F(\omega)\left(\frac{1}{\omega+z^*}-\frac{1}{\omega-z}\right)\right)=0$ everywhere on $\mathcal{C}_P$ and define the integral

$$P=\wp\int_{\mathcal{C}_P} F(\omega)\left(\frac{1}{\omega+z^*}-\frac{1}{\omega-z}\right)\mathrm{d}\omega \tag{257}$$

which has the property that

$$\mathrm{Im}\,P=0 \tag{258}$$

The contour $\mathcal{C}_P$ traverses the poles at $\omega=z$ and $\omega=-z^*$ as well as any pole at $\omega=0$, where the integral is given by the principal value.

A further contour $\mathcal{C}_B$ is defined to run from $\omega=-\infty$ to $\omega=+\infty$ but this time passing below the poles at $\omega=z$ and $\omega=-z^*$ while continuing to pass through $\omega=0$ along a segment of the real axis, where again the principal value is taken. This defines the integral

$$B=\wp\int_{\mathcal{C}_B} F(\omega)\left(\frac{1}{\omega+z^*}-\frac{1}{\omega-z}\right)\mathrm{d}\omega \tag{259}$$

where here the $\wp$ symbol denotes the principal value at $\omega=0$. The original integral is given by

$$A=\wp\int_{-\infty}^{\infty} F(\omega)\left(\frac{1}{\omega+z^*}-\frac{1}{\omega-z}\right)\mathrm{d}\omega \tag{260}$$

which is taken along the real $\omega$ axis. In terms of these integrals, the integral $P$ is given by

$$P=\tfrac{1}{2}(A+B) \tag{261}$$

Now consider the closed contour $\mathcal{C}_C$ running along the real axis from $\omega=-\infty$ to $\omega=+\infty$ and returning along $\mathcal{C}_B$. The integral around this contour is given by

$$C = \oint_{\mathcal{C}_C} F(\omega)\left(\frac{1}{\omega + z^*} - \frac{1}{\omega - z}\right)\mathrm{d}\omega$$
$$= B - A \tag{262}$$

Hence, from (261) and (262),

$$A = P - \tfrac{1}{2}C \tag{263}$$

and, since $P$ is real,

$$\operatorname{Im} A = -\tfrac{1}{2}\operatorname{Im} C \tag{264}$$

The contour $\mathcal{C}_C$ encloses the poles at $\omega = z$ and $\omega = -z^*$. It also traverses any pole that may exist at $\omega = 0$ as a principal value, but does so twice in opposite directions, so the net contribution to $C$ from $\omega = 0$ vanishes. The integral $C$ is therefore given, by Cauchy's residue theorem, as follows

$$C = 2\pi \mathrm{i}\left(-F(z) + F(-z^*)\right) \tag{265}$$

and hence, finally

$$\int_{-\infty}^{+\infty} F(\omega)\operatorname{Im}\left(\frac{1}{\omega + z^*} - \frac{1}{\omega - z}\right)\mathrm{d}\omega = \operatorname{Im} A = -\tfrac{1}{2}\operatorname{Im} C = \pi \operatorname{Re}\left[F(z) - F(-z^*)\right] \tag{266}$$

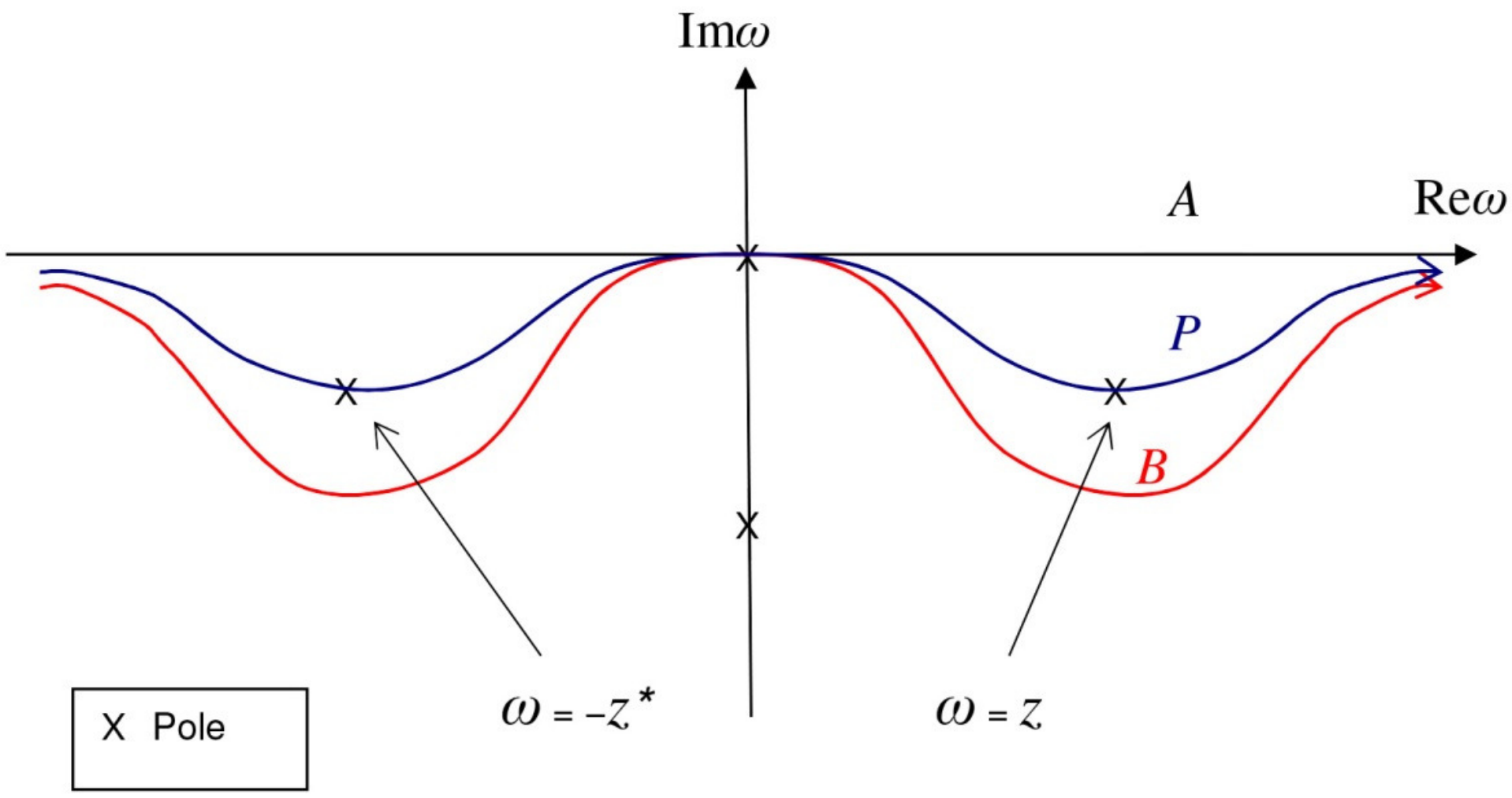

**Figure 6**: Schematic of complex-$\omega$ plane illustrating the various paths $A$, $B$ and $P$. The closed contour $C$ runs along $A$ from $\omega=-\infty$ to $\omega=+\infty$ and returns along $B$, enclosing the poles at $\omega=z$ and $\omega=-z^*$. The contour P follows $\mathrm{Im}\left(F(\omega)\left(\frac{1}{\omega+z^*}-\frac{1}{\omega-z}\right)\right)=0$ and passes through the poles at $\omega=z$ and $\omega=-z^*$. All contours pass through the pole at $\omega=0$.

## B.3 Evaluation of the integral $\int(1-\exp(-\beta\omega))^{-1}\mathrm{Im}\{1/(x^2-(\omega+\mathrm{i}y)^2)\}\mathrm{d}\omega$

The foregoing is now applied to the evaluation of the integral

$$I(x,y,\beta)=\int_{-\infty}^{+\infty}\frac{1}{1-\exp(-\beta\omega)}\mathrm{Im}\frac{1}{x^2-(\omega+\mathrm{i}\,y)^2}\mathrm{d}\omega \tag{267}$$

where $x, y$ and $\beta$ are real numbers such that for $\beta\geq 0,\ \ x>0,\ \ y>0$ and the integral is evaluated on the real-$\omega$ axis. Transforming the integral into the form considered in B.1, and applying the result (266), yields

$$\begin{aligned}I(x,y,\beta)&=\frac{1}{2x}\int_{-\infty}^{+\infty}\frac{1}{1-\exp(-\beta\omega)}\mathrm{Im}\left(\frac{1}{\omega+(x+\mathrm{i}\,y)}-\frac{1}{\omega-(x-\mathrm{i}\,y)}\right)\mathrm{d}\omega\\&=\frac{\pi}{2x}\mathrm{Re}\left(\frac{1}{1-\exp(-\beta(x-\mathrm{i}\,y))}-\frac{1}{1-\exp(\beta(x+\mathrm{i}\,y))}\right)\\&=\frac{\pi}{2x}\frac{\mathrm{e}^{2\beta x}-1}{\mathrm{e}^{2\beta x}+1-2\,\mathrm{e}^{\beta x}\cos\beta y}\\&=\frac{\pi}{2x}\coth\left(\tfrac{1}{2}\beta x\right)\left(1+\frac{\sin^2\left(\frac{1}{2}\beta y\right)}{\sinh^2\left(\frac{1}{2}\beta x\right)}\right)^{-1}\end{aligned} \tag{268}$$

It is readily confirmed that the undamped result is recovered in the limit $y\to 0$. However the function,

$$f(x,\delta)=x\tanh(x)\left(1+\frac{\sin^2\delta}{\sinh^2 x}\right) \tag{269}$$

turns out to be remarkably well approximated, for any $0\leq\delta\lesssim 1$, by

$$\begin{aligned}f(x_0,0)&=x_0\tanh(x_0)\\x_0&=\sqrt{x^2+\delta^2}\end{aligned} \tag{270}$$

for all $x\geq 0$; with a maximum numerical error (for $\delta=1$) of $\sim 8\%$. For small $x_0$,

$$f(x,\delta)\to f(x_0,0)\left(1-\tfrac{4}{45}\delta^2 x_0{}^2+\delta^4\mathcal{O}\left(x_0{}^4\right)\right) \tag{271}$$

while, for $x_0\to\infty$,

$$f(x,\delta) \sim f(x_0,0)\left(1-\frac{\delta^2}{2{x_0}^2}\right) \tag{272}$$

Therefore, in conclusion, for $0 \le \frac{1}{2}\beta y \lesssim 1$, and all $x \ge 0$, the integral (267) is given, to a good approximation, by,

$$I(x,y,\beta) \simeq \frac{\frac{1}{2}\pi}{\sqrt{x^2+y^2}}\coth\left(\tfrac{1}{2}\beta\sqrt{x^2+y^2}\right) \tag{273}$$

which becomes exact in the limit $y \to 0$.

# APPENDIX C: LIST OF SYMBOLS

## C.1 List of Symbols Used for Mathematical and Physical Quantities

Note: Throughout this article, Planck's constant, $\hbar$, and Boltzmann's constant, $k_{\mathrm{B}}$, are both set equal to unity. This means that temperature, energy and frequency are to be considered as given in the same units.

*A, B* $A_{\mathbf{q}}^{a}$, $B_{\mathbf{q}}^{a}$, coefficients in the modal decomposition (163) of the dynamical structure factor.

*a, b* General particle species labels (used in subscripts and superscripts), eg, electrons (e) and ions (i).

c Speed of light.

*D* Classical Debye length defined by (41).

$\mathfrak{D}$ Generalized screening length defined by (191).

*E* $E(\mathbf{k})$, energy of single-particle state **k**.

$E_{\mathrm{p}}$ Incident particle kinetic energy.

*e* Unit electronic charge.

*F* $F(\alpha,\beta)$, coefficient and function introduced for the GPPA at equation (102). $F^{0}(\alpha)=F(\alpha,0)$.

*G* $G(\alpha,\beta)$ function defined by (107) to be the square of the ratio of the frequency of a plasma mode to the limiting plasma frequency for that mode $=(\Omega_{\alpha\beta}/\Omega)^{2}$. $G_{\mathbf{q}}^{a}=(\Omega_{\mathbf{q}}^{a}/\Omega_{a})^{2}$; $G^{0}(\alpha)=G(\alpha,0)$; $G_{a}(\alpha,\beta)=(\Omega_{\alpha\beta}^{a}/\Omega_{a})^{2}$.

$g_{\mathbf{q}}$ $g_{\mathbf{q}}(\omega)$, function defined by (154).

$\mathfrak{g}$ $\mathfrak{g}(a|b)$, pair correlation function for particle species $a$ with respect to a particle of species *b*.

$h_{\alpha\beta}$ $h_{\alpha\beta}(\omega)$ = function defined by (104).

$I_{j}$ $I_{j}(x)$, Fermi integral as defined by (56).

**k** Single-particle momentum/wavevector.

$k_{\mathrm{F}}$ Fermi momentum = $\sqrt{2mT_{\mathrm{F}}}$

$\mathfrak{K}$ Isothermal compressibility.

*L* $L(u,v;\eta)$, function defined by (57).

*m* Particle mass.

| | |
|---|---|
| $m_e$ | Electron mass |
| $m_i$ | Ion mass. |
| $n$ | Particle number density. |
| $n$ | $n(\omega)$, refractive index. |
| $P$ | Pressure |
| $P$ | $P(z)$, the pole part of the GPPA dielectric function as defined by (121). |
| $P_\pm$ | Quantities defined by (124) - (125). |
| $\wp$ | Principal value operator. |
| $p$ | $p(\mathbf{k})$, probability, given by the Fermi-Dirac distribution (54), that the single-particle state **k** is occupied. |
| $\mathbf{q}$ | Wavevector |
| $q$ | Wavenumber. |
| $R_\mathbf{q}$ | Function defined by (43). |
| $\tilde{R}_\mathbf{q}$ | Function defined by (42). |
| $S_\mathbf{q}$ | Function defined by (44). |
| $S$ | $S(\mathbf{q},\omega)$, dynamic structure factor. |
| $S^0$ | $S^0(\mathbf{q},\omega)$, non-interacting dynamic structure factor. |
| $S$ | $S(\mathbf{q})$, static structure factor |
| $S_{ab}^{(0)}$ | $S_{ab}^{(0)}(\mathbf{q})$, quasi-classical static structure factor for particle species *a* and *b*. |
| $s$ | Parameter defined by (66) or (74). |
| $T$ | Temperature. |
| $T_F$ | Fermi temperature. |
| $t$ | Parameter defined by (74). |
| $u$ | (when used as a parameter) $= \omega/T$ |
| $u_0$ | $u_0(\mathbf{q})$, bare Coulomb potential. |
| $u$ | $u(\mathbf{q},\omega)$, dynamically screened Coulomb potential. |

$u_{\alpha\beta}$ $= \Omega_{\alpha\beta}/T$

$\mathfrak{V}$ Volume.

$v$ Parameter defined by $v = q^2/2mT$ .

$\boldsymbol{v}$ Velocity.

$w_a$ Particle species charge weighting factor, defined generally by (176).

$Z$ Ion charge.

$z$ Complex frequency parameter, as defined by (106), or a general complex variable.

$\alpha$ Static screening parameter. In particular: $\alpha_{\mathbf{q}} = \varepsilon(\mathbf{q},0)-1$ ; $\alpha_a = \alpha_{\mathbf{q}}^a = \varepsilon_a(\mathbf{q},0)-1$ .

$\alpha$ In context of a multicomponent plasma (section 4.2 only) $= \sum_a \alpha_a$

$\beta$ GPPA mode displacement parameter. (Represents the shift in the modal parameters from their OCP values to those applicable to a multicomponent plasma.)

$\Gamma(z)$ Euler gamma function.

$\gamma$ $\gamma(\eta)$ , coefficient in the formula (79).

$\Delta_{ab}$ Coefficient defined at equation (201).

$\delta(x)$ Dirac delta function.

$\varepsilon_0$ Permittivity of free space.

$\varepsilon$ $\varepsilon(\mathbf{q},\omega)$ , dielectric function.

$\varepsilon_{\pm}$ Solutions of the GPPA equation (120) for the OCP dielectric function.

$\eta$ Degeneracy parameter, $=\mu/T$ .

$\mathrm{K}_{\mathbf{q}}$ $\mathrm{K}_{\mathbf{q}}(\omega)$ , Green function, which gives the charge-density response to an applied electric field.

$\kappa$ $\kappa(\omega)$ , absorption opacity; $\rho\kappa$ = absorption coefficient.

$\lambda$ Charge scaling parameter in (22).

$\lambda_a$ Coefficient in the GPPA multicomponent expansion, (136).

$\lambda_{ab}$ Coefficient satisfying (198).

$\mu$ Chemical potential.

$\nu_{\mathbf{q}}$ Collision or mode damping frequency associated with plasma collective mode **q**, $\nu_{\mathbf{q}} = -2\,\mathrm{Im}\,\widehat{\Omega}_{\mathbf{q}}$.

$\nu_{\alpha\beta}$ Damping or effective collision frequency for a multicomponent plasma expressed in terms of the GPPA parameters, $\alpha, \beta$. Equivalent to $\nu_{\mathbf{q}}$.

$\Pi_{\mathbf{q}}$ $\Pi_{\mathbf{q}}(\omega)$, polarization part of the response Green function.

$\rho$ Mass density.

$\sigma$ $\sigma(\omega)$, electrical conductivity.

$\sigma_{\mathbf{q}}$ $\sigma_{\mathbf{q}}(\omega)$, spectral density, (3) - (6).

$\Phi(x)$ Dawson's integral, as defined by (71).

$\Omega$ (Limiting) plasma frequency, given by (15).

$\Omega_{\mathrm{i}}$ Ion plasma frequency $= \sqrt{Z^2 e^2 n_{\mathrm{i}} / \varepsilon_0 m_{\mathrm{i}}}$

$\Omega_{\mathbf{q}}$ Effective frequency of plasma collective mode **q**, where $\left(\Omega_{\mathbf{q}}\right)^2 = \left(\hat{\Omega}_{\mathbf{q}}\right)^2 + \frac{1}{4}\left(\nu_{\mathbf{q}}\right)^2$

$\widehat{\Omega}_{\mathbf{q}}$ Complex plasma mode frequency, defined by $\varepsilon\left(\mathbf{q}, \widehat{\Omega}_{\mathbf{q}}\right) = 0$.

$\hat{\Omega}_{\mathbf{q}}$ $= \mathrm{Re}\,\widehat{\Omega}_{\mathbf{q}}$.

$\Omega_{\alpha\beta}$ Collective mode frequency, $\Omega_{\mathbf{q}}$, of a multicomponent plasma, as represented in terms of the GPPA parameters, $\alpha, \beta$.

$\omega$ Frequency.

This page is intentionally left blank